\documentclass[aps,preprint,groupedaddress,nofootinbib]{revtex4}
\usepackage{amsmath}
\usepackage{amsfonts}
\usepackage{graphicx}

\def\dalemb#1#2{{\vbox{\hrule height.#2pt
       \hbox{\vrule width.#2pt height#1pt \kern#1pt \vrule width.#2pt}
       \hrule height.#2pt}}}

\def\ba{\begin{eqnarray}}
\def\ea{\end{eqnarray}}
\def\be{\begin{equation}}
\def\ee{\end{equation}}

\def\gtorder{\mathrel{\raise.3ex\hbox{$>$}\mkern-14mu
            \lower0.6ex\hbox{$\sim$}}}
\def\ltorder{\mathrel{\raise.3ex\hbox{$<$}\mkern-14mu
            \lower0.6ex\hbox{$\sim$}}}

\def\ellb{\boldsymbol{\ell }}
\def\thetab{\boldsymbol{\theta }}

\begin{document}

\title{CMB Lensing Reconstruction in Real Space}

\author{Martin Bucher${}^{1,2,4}$ 
Carla Sofia Carvalho${}^{3}\footnote{Corresponding author:
carvalho.c@gmail.com},$ 
Kavilan Moodley${}^{3,4}$ 
and Mathieu Remazeilles${}^{1,2}$ 
}

\affiliation{${}^{1}$Laboratoire APC, 
Universit\'e Paris Diderot-Paris 7,
B\^atiment Condorcet,
Case 7020,
75205 Paris Cedex 13, France}

\affiliation{${}^{2}$Laboratoire de Physique Th\'eorique, 
B\^atiment 210, Universit\'e
Paris-Sud, 91405 Orsay Cedex, France}

\affiliation{${}^{3}$
Astrophysics and Cosmology Research Unit, 
University of KwaZulu-Natal, Westville, Durban 4000, South Africa
}

\affiliation{${}^4$
National Institute for Theoretical Physics (NITheP), Private Bag X54001, Durban 4000, South Africa.
}

\begin{abstract}
We explore the reconstruction of the gravitational
lensing field of the cosmic microwave background in real space showing
that very little statistical information is lost when estimators
of short range on the celestial sphere are used in place of the 
customary estimators
in harmonic space, which are nonlocal and in principle require
a simultaneous analysis of the entire sky without any cuts or excisions.
Because virtually all the information relevant to lensing reconstruction
lies on angular scales close to the resolution scale of the sky map, 
the gravitational lensing dilatation and shear fields (which unlike
the deflection field or lensing potential are directly related to
the observations in a local manner) may be reconstructed by means of
quadratic combinations involving only very closely separated 
pixels. Even though harmonic space provides a more natural context for 
understanding lensing reconstruction theoretically, the real space 
methods developed here have the virtue of being faster to implement and 
are likely to prove useful for analyzing 
realistic maps containing a galactic cut and possibly numerous small
excisions to exclude point sources that cannot be reliably subtracted. 
\end{abstract}


\author{}
\affiliation{}

\date{\today}


\maketitle

\section{Introduction}

Recently there has been much discussion of how
the intervening mass from clustered matter between
the last scattering surface (at $z\approx 1100$)
and an observer at the present time distorts the 
CMB anisotropy by means of gravitational lensing \cite{blanchard,cole,tomita,bernardeau,van-waerbeke}.
On the level of the two-point correlation function, 
this effect distorts the TT (temperature-temperature)
correlation power spectrum \cite{seljak} and 
mixes the EE and BB 
polarization power spectra as well as distorting them \cite{zald,benabed-bis,seljak-ter}. 
Lensing also introduces 
non-Gaussianities that manifest themselves in the higher-point correlation
functions \cite{zaldarriaga,cooray-ter}. At the level of the three-point correlation
function, to leading order there is no nonzero expectation
value if we regard the lensing potential as a random field \cite{cooray,kesden}.
However if we consider the CMB lensing potential as fixed,
we find that expectation values of the form 
\ba 
\left< 
T (\boldsymbol{\theta})
T(\boldsymbol{\theta}')
\right> _{\Phi (\boldsymbol{\theta}^{\prime \prime })}
\ea
do not vanish, and this property may be exploited to
recover or ``reconstruct'' the lensing field
using estimators quadratic in $T$ (or in $E$ and $B$) \cite{hu2001a}.

Much effort has been devoted to developing an optimal 
reconstruction of the lensing potential in harmonic space, 
which implicitly assumes full sky
coverage with no galactic cut, no bad pixels due to
point sources that must be excised, and no nonuniform
weighting to account for uneven sky coverage \cite{challinor,lesgourgues,bode}. 
For a reconstruction based on the temperature anisotropy
alone, it has been shown how to construct the optimal quadratic
estimator in this idealized context \cite{hu2001b}, and the improvement
that can be gained from using an even more optimal
maximum likelihood estimator is marginal \cite{seljak-bis}, because the 
distortion due to lensing is small compared to the 
intrinsic cosmic variance and noise of the experiment
(although this assumption is less valid at very
large $\ell $ for very clean maps where the lensing 
signal is dominant). For the exploitation
of polarized anisotropies, the situation is somewhat
more complicated. When the experimental noise is large
compared to the B polarization mode generated by lensing,
the situation is essentially the same as for a reconstruction using
the temperature data \cite{hu-okamoto,cooray-bis}.
However at a higher sensitivity where the B signal
is essentially entirely due to lensing, the quadratic estimator
underperforms because the actual multipole moments rather
than their averages should be used for the optimal weighting. 
In this case, the higher order corrections to the 
quadratic estimator present in the maximum likelihood
estimator are no longer negligible \cite{hirata-seljak}.  

In this paper we investigate a real space approach
to lensing reconstruction. Under the ideal conditions
often assumed and described above, this approach would naturally
yield the same result as the conventional approach in harmonic space.
Our interest however lies in considering slightly non-optimal
estimators that have been modified to have a finite range so that cuts, 
excisions of pixels, and non-uniform coverage may be included in a simple
and flexible way. We believe that
such non-ideal but more robust local estimators
defined in real space may prove superior for confronting the complications
inherent in analyzing real data \cite{smith,miller,perotto}. 
Another advantage of the approach here is that the dilatation and pure shear
provide separate and essentially independent lensing reconstructions which
may be confronted with each other. This feature may prove useful
as a way of diagnosing spurious signals, which are unlikely to affect
the two reconstructions in the same way. Moreover the presence of
two shear components enables one to estimate the noise of the 
reconstruction through the implied transverse displacement field, 
which is forbidden in weak lensing. 

Before proceeding to the details of this program, it is useful
to consider the relation between the various descriptions
of the lensing and the deformation of the anisotropies in real
space. It is also useful to consider which angular scales 
contribute the most statistical weight to the lensing reconstruction.

The lensing distortion of the CMB anisotropy on
the surface of last scatter may be described in three ways:
by a lensing potential $\Phi ,$ by a deflection field 
$\boldsymbol{\xi }=\nabla \Phi ,$ or by the three components of
the shear tensor
\ba 
\kappa =
\begin{pmatrix}
\kappa _0+\kappa _{+}& \kappa _{\times }\cr
\kappa _{\times }& \kappa _0-\kappa _{+}\cr
\end{pmatrix}
=\nabla _a\nabla _b\Phi .
\ea 
Even if we have simultaneous access to the entire sky, 
the descriptions $\Phi $ and $\boldsymbol{\xi }$ suffer
from an ambiguity. $\Phi $ cannot be distinguished 
from $\Phi +\textrm{(constant)}$ and the vector field
$\boldsymbol{\xi }$ can be measured only up to a 
constant translation (or more properly a rotation in the 
presence of sky curvature). 
This is because if we know only the CMB power spectrum,
a patch of sky and its translation necessarily have
the same likelihood on account of isotropy of the
underlying stochastic process. Consequently,
the absolute translation due to lensing cannot be observed.
By contrast, locally the shear and dilatation (which are gradients
of the translation vector field) are completely well defined.
This can easily be seen by considering 
the effect of a constant deformation described by 
a deformation matrix $S$ relating the angular coordinates
$\boldsymbol{\theta },$ the actual coordinates on the celestial
sphere of a point on the last scattering surface, to
the coordinates $\boldsymbol{\theta }'$, the 
coordinates that the same point would have in the 
absence of lensing. 
\footnote{In the sequel we shall, 
unless otherwise indicated, employ the flat sky approximation
where the vector ${\thetab }$ represents a point
on the flattened celestial sphere and ${\ellb }$ represents
a wavevector. At times summations over $(l,m)$ shall
also be used.}
We have 
$\boldsymbol{\theta '}= S\boldsymbol{\theta }$ 
where $S=\exp [-\boldsymbol{\kappa }].$
(Note that we employ the flat sky approximation
and assume that the deformation is small so that a linear
treatment is adequate.)
To linear order, the power spectrum is modified
in the following way by this linear deformation,
which preserves the homogeneity but not the isotropy
of the underlying statistical process:
\ba 
C(\vert {\ellb }\vert )
\to C(\ell ) \biggl[
1 + \kappa _0
\left( \frac{d(\ln[C(\ell )])}{d(\ln [\ell ])} + 2\right)
+\left( 
\frac{\kappa _+(\ell _1^2-\ell _2^2)+\kappa _\times  (2\ell _1\ell _2)}{\ell ^2}
\right)
\frac{d(\ln [C(\ell )])}{d(\ln [\ell ])}
\biggr] .
\ea 
For the case of perfect scale invariance 
(i.e., a power law of the form $C(\ell )\propto \ell ^{-2}$) 
there is no change in the 
correlations due to the dilatation component of $S,$
and similarly for a perfect white noise spectrum
(i.e., a power law of the form $C(\ell )\propto \ell ^0$)
there is no sensitivity to the pure shear components
$\kappa _+$ and $\kappa _\times $ in the anisotropic
$(m=\pm 2)$ correlations.  

To the extent that the shear-dilatation components
are slowly varying, we may construct estimators 
of $\kappa _0,$ $\kappa _+,$ and $\kappa _\times $ as 
follows
\ba 
\hat \kappa _0 &=& {N_0}^{-1}
\int _A d^2\theta ~
\int _A d^2\theta '~
\left[ 
T(\thetab )~T(\thetab ')-
\left< T({\thetab })~T({\thetab '})\right> _{\kappa =0}
\right]
\cr
&&\times \int \frac{d^2\ell }{(2\pi )^2}
\exp [i{\ellb }\cdot ({\thetab }-{\thetab '})]~
\frac{C(\ell )}{[C(\ell )+N(\ell )]^2}
\left( \frac{d(\ln [C(\ell )])}{d(\ln [\ell ])} + 2\right)\cr 
&=& {N_0}^{-1}{\cal A}^{-1}
\int _A d^2\theta ~
\int _A d^2\theta '~
\left[ 
T({\thetab })~T({\thetab '})-
\left< T({\thetab })~T({\thetab '})\right> _{\kappa =0}
\right]
K_0({\thetab }-{\thetab '})\cr
&=& \frac{1}{\cal A}
\int _A d^2\theta ~
\left[
T({\thetab })~({\cal F}_{\kappa _0}T)({\thetab }) -\textrm{(constant)}
\right]
\ea
and similarly
\ba 
\begin{pmatrix}
\hat \kappa _{+}\cr 
\hat \kappa _{\times }\cr 
\end{pmatrix}
&=& {N_{+,\times }}^{-1}
\int _A d^2\theta ~
\int _A d^2\theta '~T(\thetab )~T({\thetab '})~\cr
&&\times \int \frac{d^2\ell }{(2\pi )^2}
\exp [i{\ellb }\cdot ({\thetab }-{\thetab '})]~
\frac{C(\ell )}{[C(\ell )+N(\ell )]^2}
\frac{d(\ln [C(\ell )])}{d(\ln [\ell ])}
\begin{pmatrix}
\cos (2\vartheta )\cr
\sin (2\vartheta )\cr
\end{pmatrix}\cr 
&=& {N_{+,\times }}^{-1}{\cal A}^{-1}
\int _A d^2\theta ~
\int _A d^2\theta '~T({\thetab })~T({\thetab '})~
\begin{pmatrix}
K_+({\thetab }-{\thetab '})\cr
K_\times ({\thetab }-{\thetab '})\cr 
\end{pmatrix},
\ea 
where the normalization factors are given by 
\ba 
N_0&=&{\cal A}
\int \frac{d^2\ell }{(2\pi )^2}
\frac{[C(\ell )]^2}{[C(\ell )+N(\ell )]^2}
\left( \frac{d(\ln [C(\ell )])}{d(\ln [\ell ])} + 2\right)^2,\cr 
N_+&=&N_\times =\frac{\cal A}{2}
\int \frac{d^2\ell }{(2\pi )^2}
\frac{[C(\ell )]^2}{[C(\ell )+N(\ell )]^2}
\left( \frac{d(\ln [C(\ell )])}{d(\ln [\ell ])}\right)^2
\ea
and ${\cal A}$ is the area of the domain. 
Here $N(\ell )$ is the noise of the experiment being considered
and serves as a cut-off at large $\ell ,$ above which there is
very little exploitable information because of the low
signal-to-noise of the CMB maps. 

The above expressions assume a spatially flat, two-dimensional 
domain formally of large but 
finite area subject to a spatially uniform linear deformation.
We may consider a toroidal domain in the limit that the 
periods of the torus become arbitrarily large (though the toroidal domain is not essential for the application of the real space estimator presented in this paper). 
Before proceeding it is useful to consider the relation of this 
simplified problem to the real problem 
of reconstructing the lensing field starting from a 
CMB map of finite resolution. Because of the smallness 
of the lensing distortion of the CMB anisotropy, 
the idealized situation considered above is 
less different from 
the actual
situation, where one is dealing with a curved sky 
and a varying dilatation and shear fields,
than one might at first sight suppose.

\begin{figure}
 \begin{center}
    \includegraphics[width=17cm]{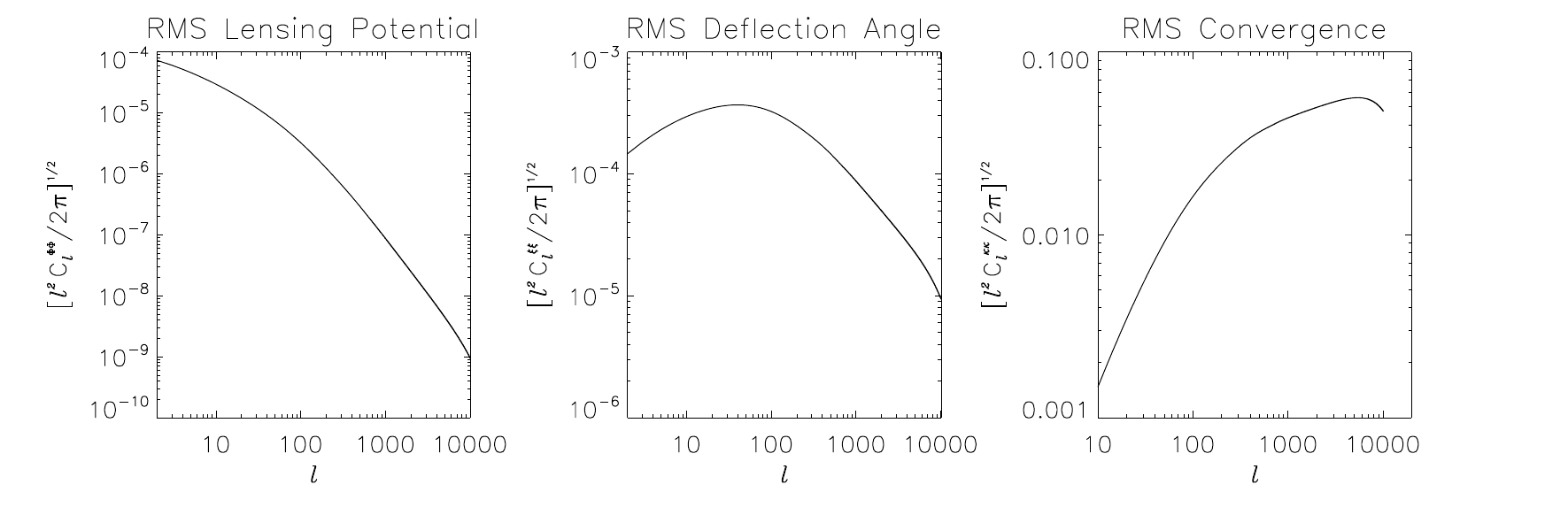}
 \end{center}
\caption{\baselineskip=0.5cm{
{\bf Lensing power spectrum.}
The lensing field power spectrum is shown represented
in several manners. The three panels (from left to right) illustrate
the lensing field expressed as potential, a deflection field,
and dilatation/shear field, respectively. Plotted
are $[\ell (\ell +1)C_{XX}/(2\pi )]^{1/2}$
where $XX=\Phi \Phi ,$ 
${\boldsymbol{\xi \xi }},$ $\kappa \kappa .$
Here
$C_{\ell }^{\boldsymbol{\xi \xi }}=\ell (\ell +1) ~C_{\ell }^{\Phi \Phi },$
and 
$C_{\ell }^{\kappa \kappa }=\ell ^2(\ell +1)^2 ~C_{\ell }^{\Phi \Phi }/4.$
Of the three, the last power spectrum is more directly related
to the observed distortion. 
}}
\label{Fig:LensingFields}
\end{figure}

The rightmost panel of Fig.~\ref{Fig:LensingFields} shows
the shear-dilatation power spectrum as a function of multipole number $\ell ,$
and we observe that for $\ell <100,$ the distortion is always
less than about $1.5\% .$ This implies that in order to attain 
an $(S/N)$ of approximately unity it is necessary to consider 
a region containing at least $O(10^3)$ resolution elements,
where a resolution element is a pixel of the map of sufficient size
so that the noise and angular resolution of the survey give $S/N\approx 1.$
Consequently, there is little point to trying to
reconstruct the lensing field
over a region not having at least 30 resolution elements on a side. 
If the distortion from lensing were greater, the situation would be 
different. 

For the ideal linear estimator 
\ba 
\frac{1}{\sigma ^2_{\hat \kappa _{0,ideal}}}=
\left(
\frac{S}{N}
\right) ^2
&=&
\sum _{\ell , m}
\frac{C_\ell ^2}
{2(C_\ell +N_\ell )^2}
\left[
\frac{d(\ln [C])}{d(\ln [\ell ])}+2
\right] ^2\cr 
&\approx &
\frac{{\cal A}}{(4\pi )}
\int _0^\infty d\ell ~\ell ~
\frac{C(\ell )^2}
{(C(\ell )+N(\ell ))^2}
\left[
\frac{d(\ln [C(\ell )])}{d(\ln [\ell ])}+2
\right] ^2
\label{SoverN}
\ea
where ${\cal A}$ is the area of the sky patch considered.

We now consider other unbiased estimators 
of $\kappa _0$ and $\kappa_+$ where a slight increase in the variance
is compensated for by other desirable properties. We
are presently interested in estimators for which the 
real space filtering kernel is short range. 
To this end it is useful to define the inner product
on the space of weight vectors ${w}=\{ w_\ell \}$:
\ba
\left< w^A, w^B \right> 
&=&
\sum _{\ell , m}
\frac{1}{2(C_\ell +N_\ell )^2}
w^A_{\ell ,m}
w^B_{\ell ,m}\cr
&=&
{\cal A}\int \frac{d^2\ell }{(2\pi )^2}
\frac{1}{2[C(\ell )+N(\ell )]^2}
w^A(\ellb )
w^B(\ellb)
\ea
where we give both the spherical and flat sky continuum
forms. 
If we set $w_{ideal}(\ell )=\delta C(\ell )_{th, \kappa _0=1}=
C(\ell)[\frac{d \ln C_\ell}{d \ln \ell}+2],$
then in terms of the above inner product
\ba
\hat \kappa _{0,ideal}= 
\frac{
\left< w_{ideal}, \delta C_{obs}\right> 
}{
\left< w_{ideal},  w_{ideal} \right> 
}
\ea
and 
\ba 
\left( \frac{S}{N}\right) ^2_{\kappa _0, ideal}=
\left<\delta C_{th, \kappa _0=1}, \delta C_{th, \kappa _0=1}\right> . 
\ea
Given an arbitrary weight vector $w,$ using the above inner
product we may define the following unbiased estimator of
$\kappa _0$
\ba 
\hat \kappa _0(w)=
\frac{
\left< w, \delta C_{obs}\right> 
}{
\left< w,  w_{ideal} \right> 
}
\ea
provided that the denominator does not vanish, and its
variance is given by
$\left< w,  w\right> /\left< w_{ideal},  w\right> ^2,$
so that the increase in variance with respect to the ideal estimator
is given by following geometric expression
for the secant squared
\ba
\frac{
\textrm{Var}(\hat \kappa _0(w))
}{
\textrm{Var}(\hat \kappa _0(w_{ideal}))
}
=
\frac{
\left< w,  w\right> 
\left< w_{ideal},  w_{ideal} \right> 
}{
\left< w,  w_{ideal} \right> ^2
}~\rm{sec}^2(\chi).
\label{VVar}
\ea
For $\kappa _+$ and $\kappa _\times $ analogous formulae may be 
derived straightforwardly.

\begin{figure}
 \begin{center}
    \includegraphics[width=7.5cm]{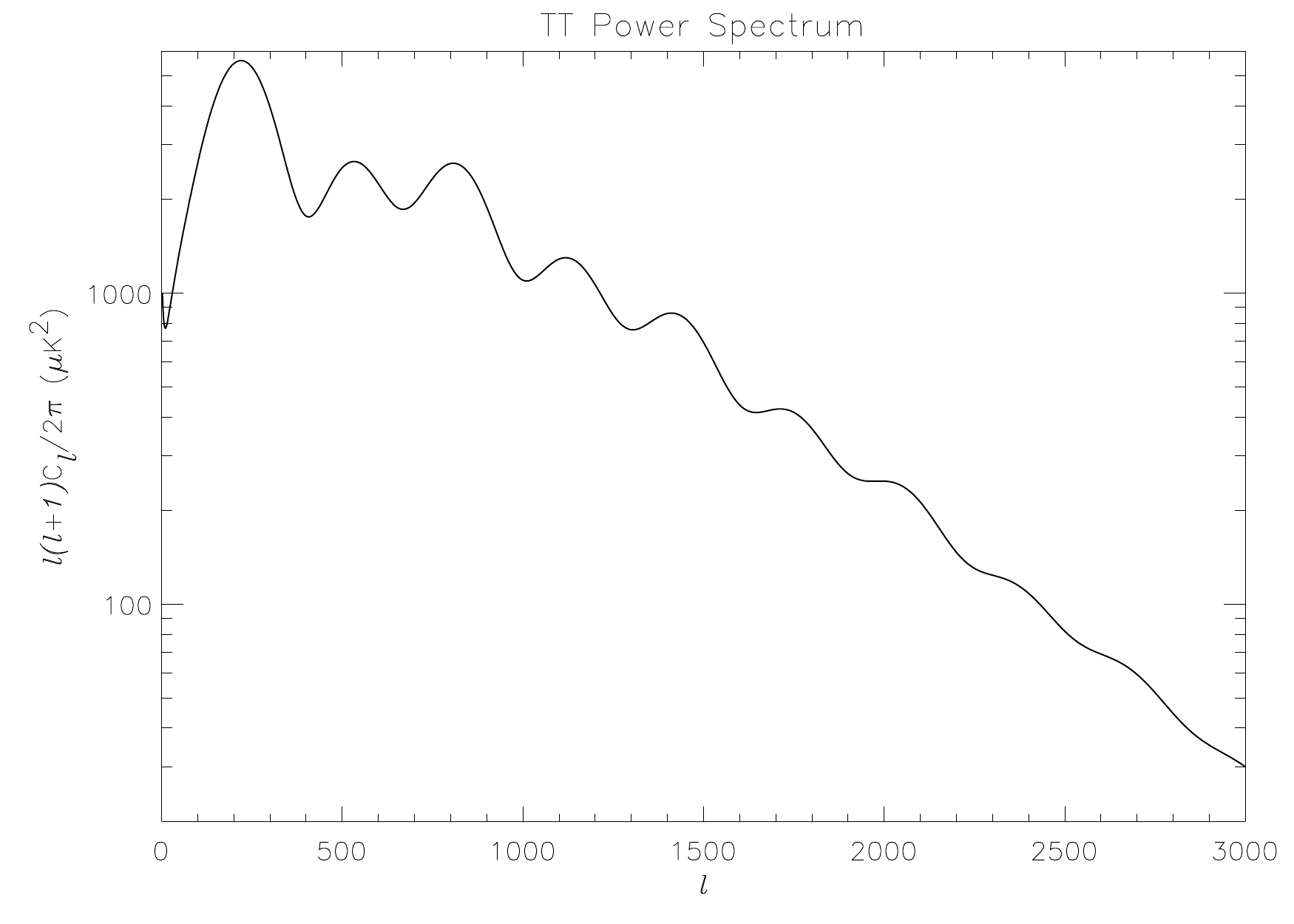}
    \includegraphics[width=7.5cm]{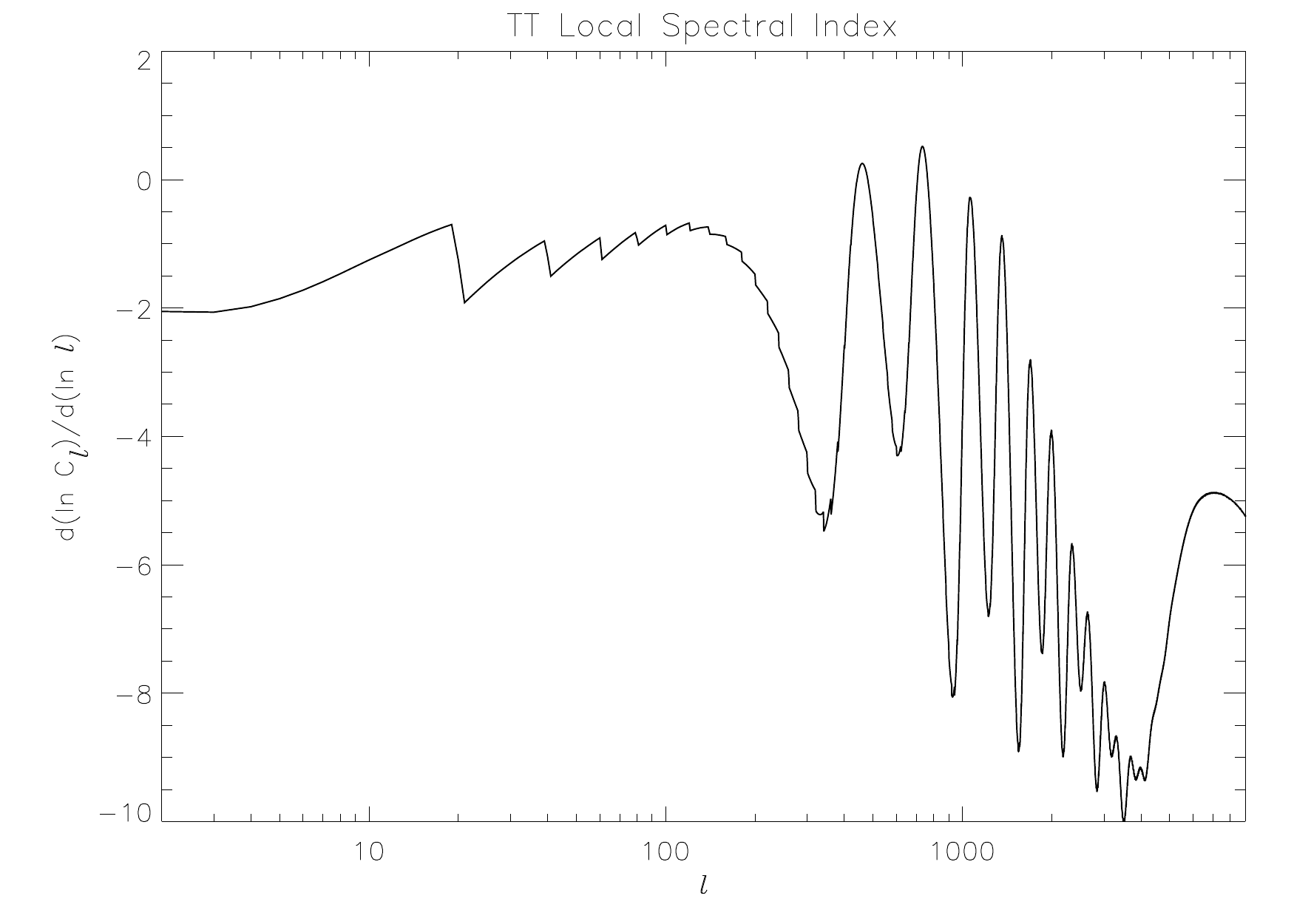}\\
    \includegraphics[width=7.5cm]{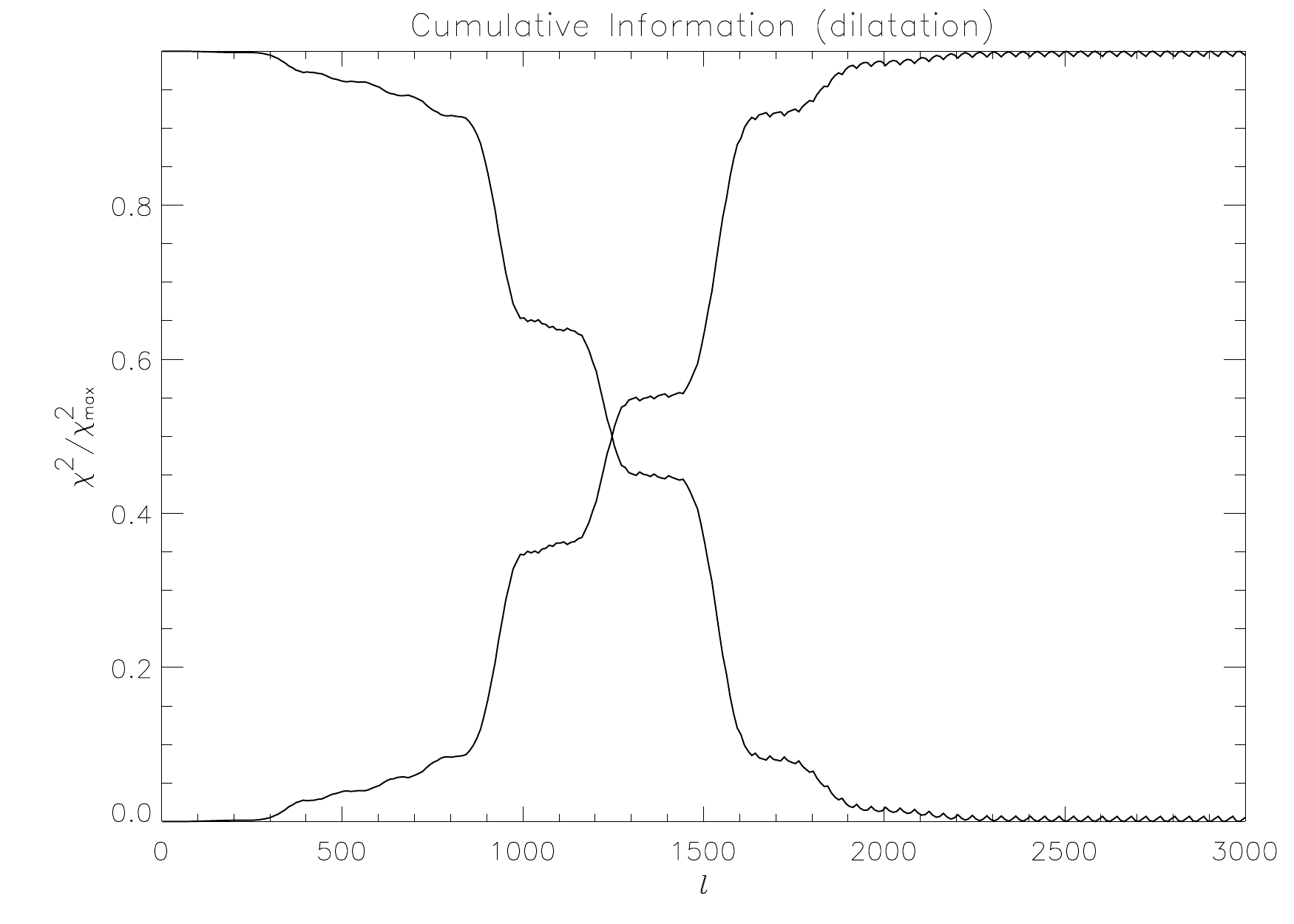}
    \includegraphics[width=7.5cm]{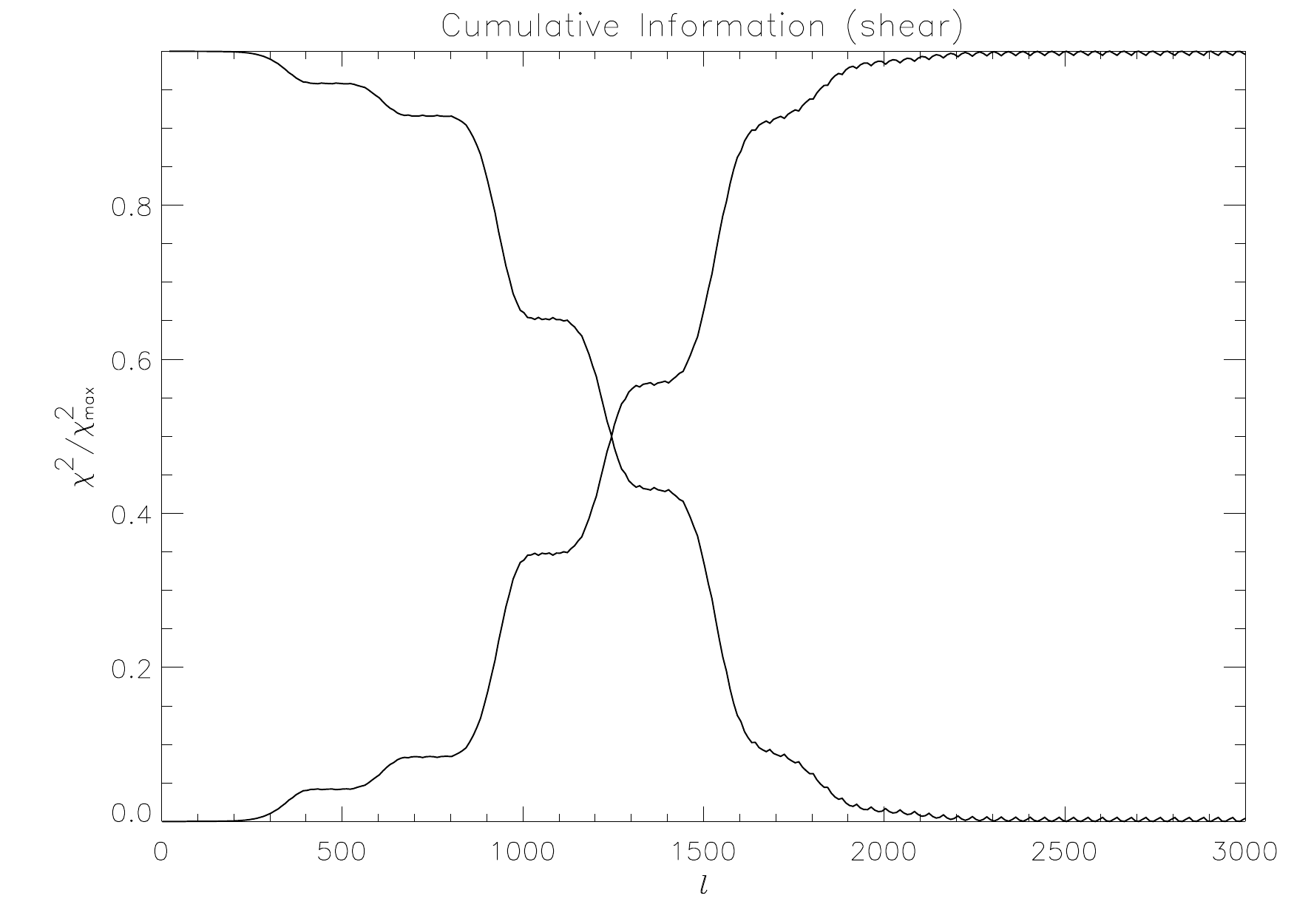}
 \end{center}
\caption{\baselineskip=0.5cm{
{\bf Character of the signal.}
The top row shows as a function of multipole number $\ell $
the temperature power spectrum and local spectral index,
defined as $d\ln C_\ell /d\ln \ell,$  
for the standard cosmology (WMAP best-fit model).
On the bottom row, 
the left panel shows the normalized cumulative 
$\chi ^2$ as a function of $\ell $ integrated both 
from the left and from the right using the sensitivity 
and resolution parameters for the PLANCK experiment
(where the 100, 143 and 217 GHz channels have been
combined in quadrature) \cite{bluebook}. We observe that the central 80\% of
the information is concentrated in the range $\ell =800$--$1600.$
Smaller $\ell $ contribute almost no information because there
are comparatively very few
independent multipoles, and moreover the angular spectrum in 
the sky is very nearly scale invariant. At much larger $\ell $
instrument noise and beam smearing wash out the usable signal.
In the intermediate range a structure of plateaus connected
by steep rises can be observed. This structure is a direct
result of the Doppler oscillations. Around the crests and troughs
the spectrum is almost scale invariant and hence does not contain any 
information for determining the dilatation. The right panel
shows the corresponding plot for the shear, where the plateaus
are less pronounced.  
}}
\label{Fig:Info}
\end{figure}

We consider how the information (or $S^2/N^2$) 
contained in the ideal estimator
is distributed over the various multipoles. In Fig.~\ref{Fig:Info}
[panels (c) and (d)]
we plot the quantity
\ba 
F_<(\bar \ell )=
\frac{
\int _0^{\bar \ell }\ell ~d\ell ~
\frac{ {C(\ell) }^2}{({C(\ell)}+{N(\ell )})^2}
\left[ \frac{d(\ln [C(\ell )])}{d(\ln [\ell ])}+2 \right] ^2
}{
\int _0^{\infty }\ell ~d\ell ~
\frac{ {C(\ell )}^2}{({C(\ell )}+{N(\ell )})^2}
\left[ \frac{d(\ln [C(\ell )])}{d(\ln [\ell ])}+2 \right] ^2
}
\ea
and $F_>(\bar \ell )=1-F_<(\bar \ell )$ where 
\ba 
N_\ell 
=N_0 ~\exp \left[ +\ell ^2\theta ^2_{beam}\right]
=N_0 ~\exp \left[ +\ell ^2/\ell ^2_{beam}\right]
\ea
and $\ell _{beam}=(810)(10'/\theta _{beam}^{fwhm}).$
The corresponding quantity is also shown for the shear. 

In their present state, the ideal minimum variance kernels for the 
estimators 
$\hat \kappa _0,$
$\hat \kappa _+,$ 
and 
$\hat \kappa _\times $ 
have their support sharply peaked at small separations, but
nevertheless there is still some small support 
for large separation. We now investigate how much information
is lost if the support at large separation is completely
cut away. The quantitative way to characterize this loss
is to ask by what factor the variance of the estimator is 
increased relative to the optimal estimator after our
pruned estimator has been renormalized to render it unbiased.

\def\half{\frac{1}{2}}

\begin{figure}
 \begin{center}
    \includegraphics[width=5.3cm]{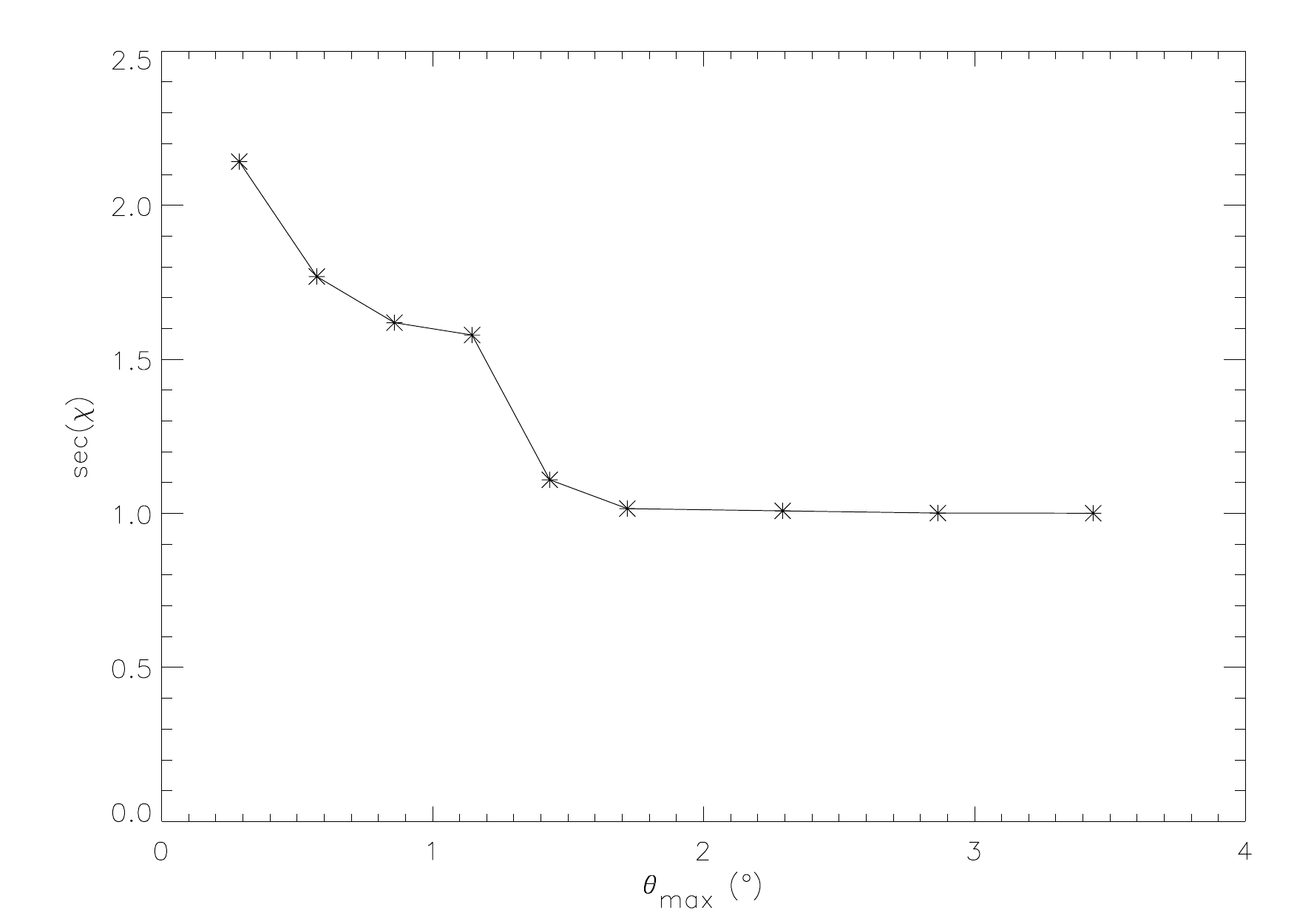}
    \includegraphics[width=5.3cm]{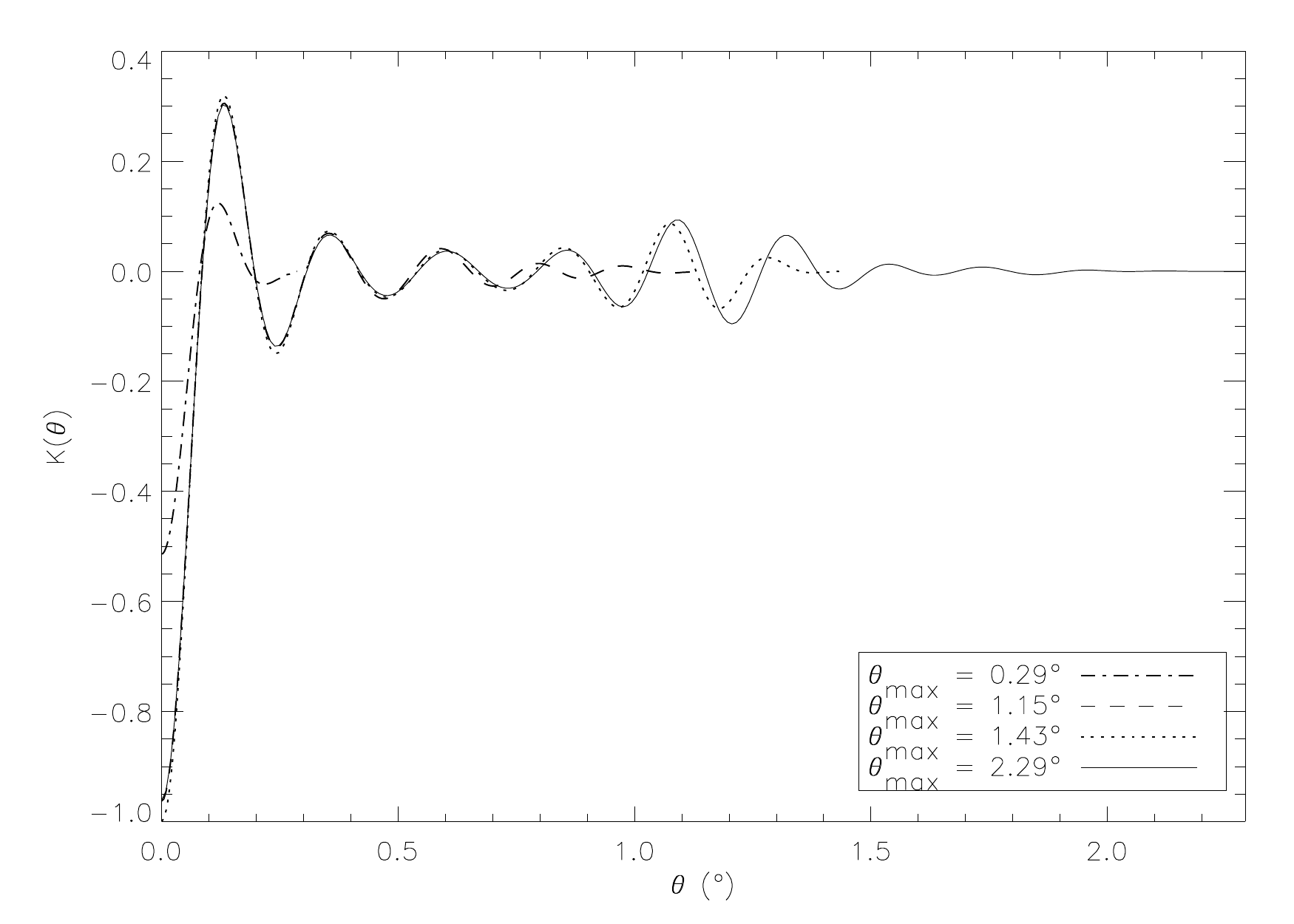}
    \includegraphics[width=5.3cm]{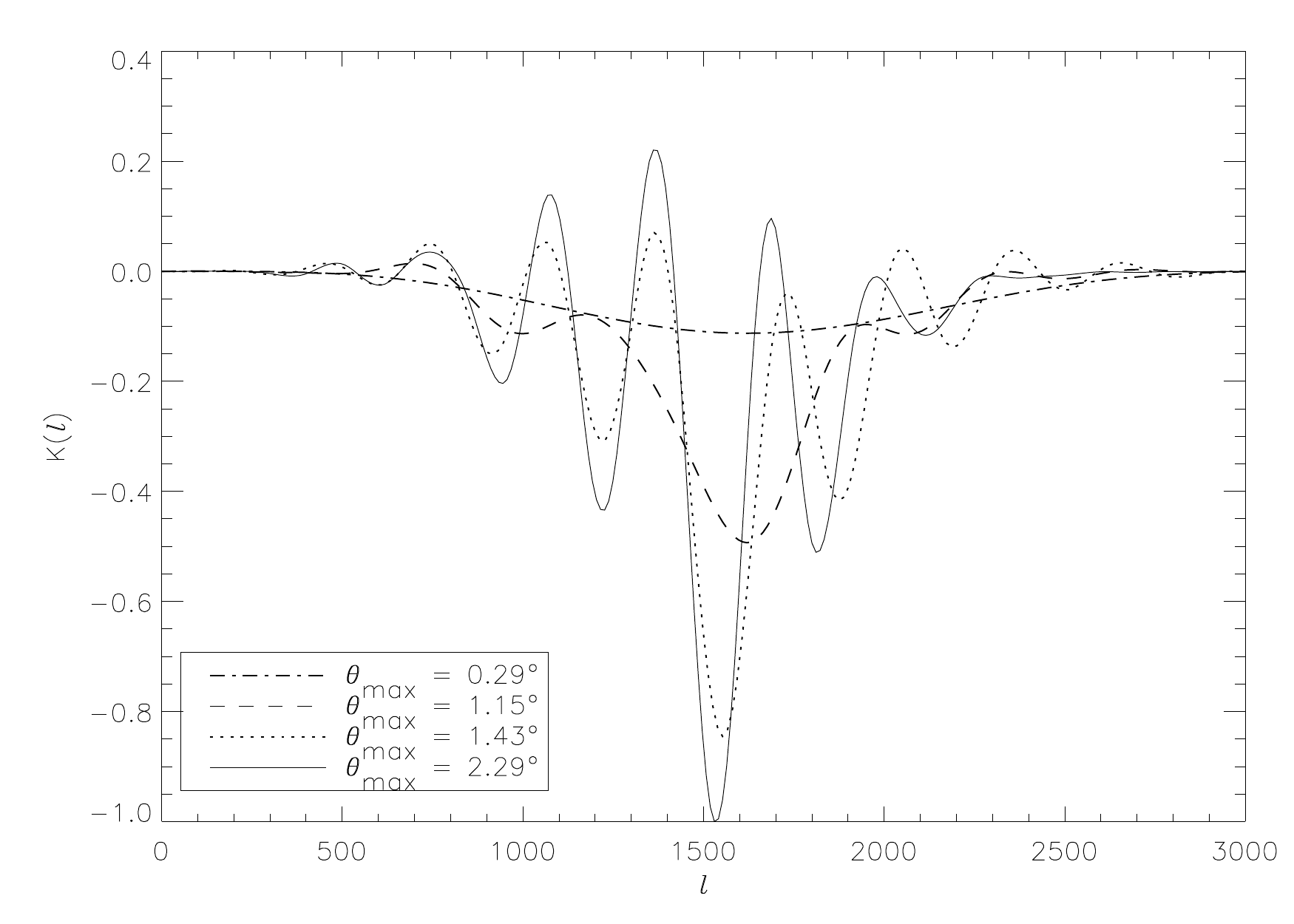}\\
    \vspace{0.5cm}
    \includegraphics[width=5.3cm]{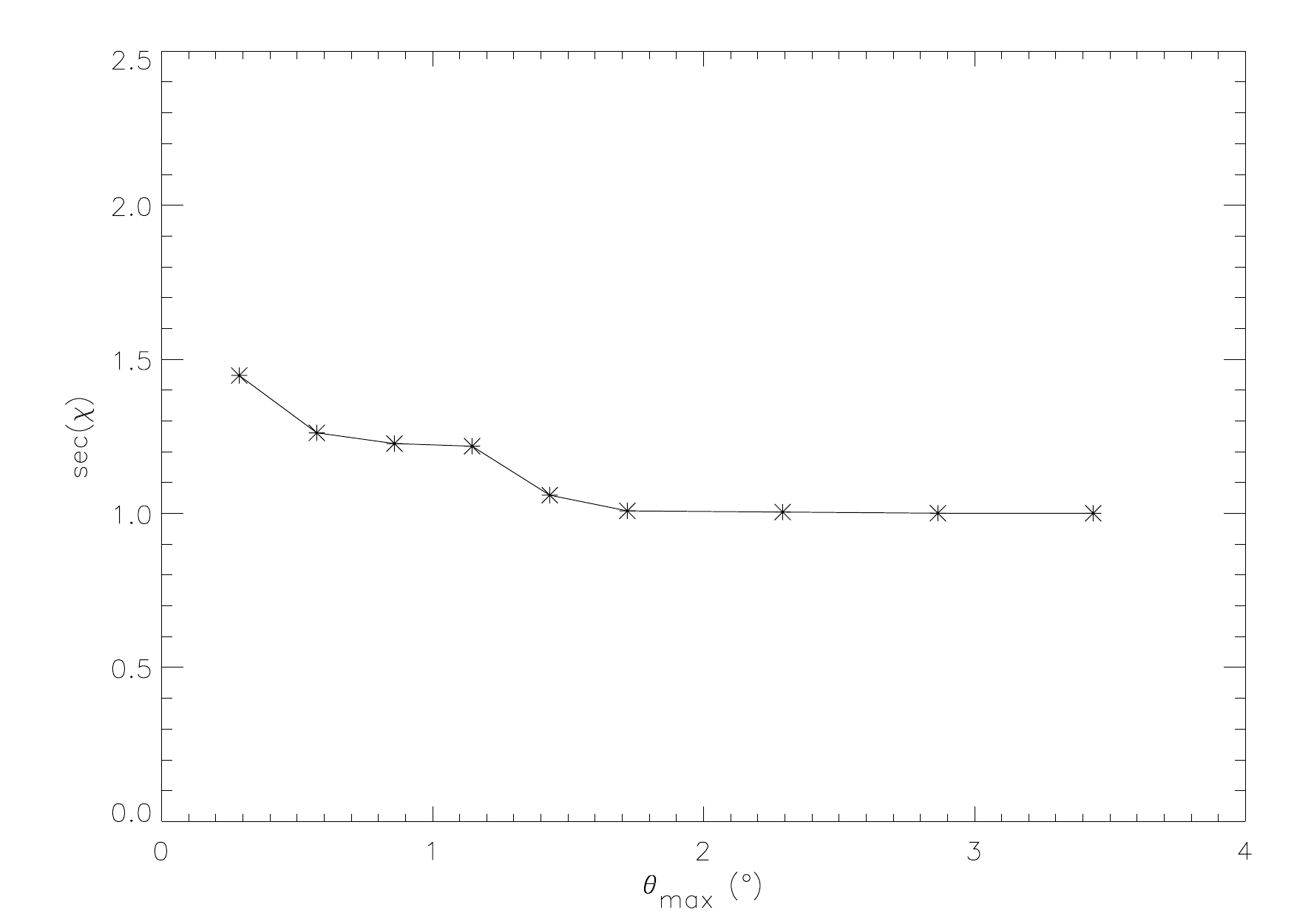}
    \includegraphics[width=5.3cm]{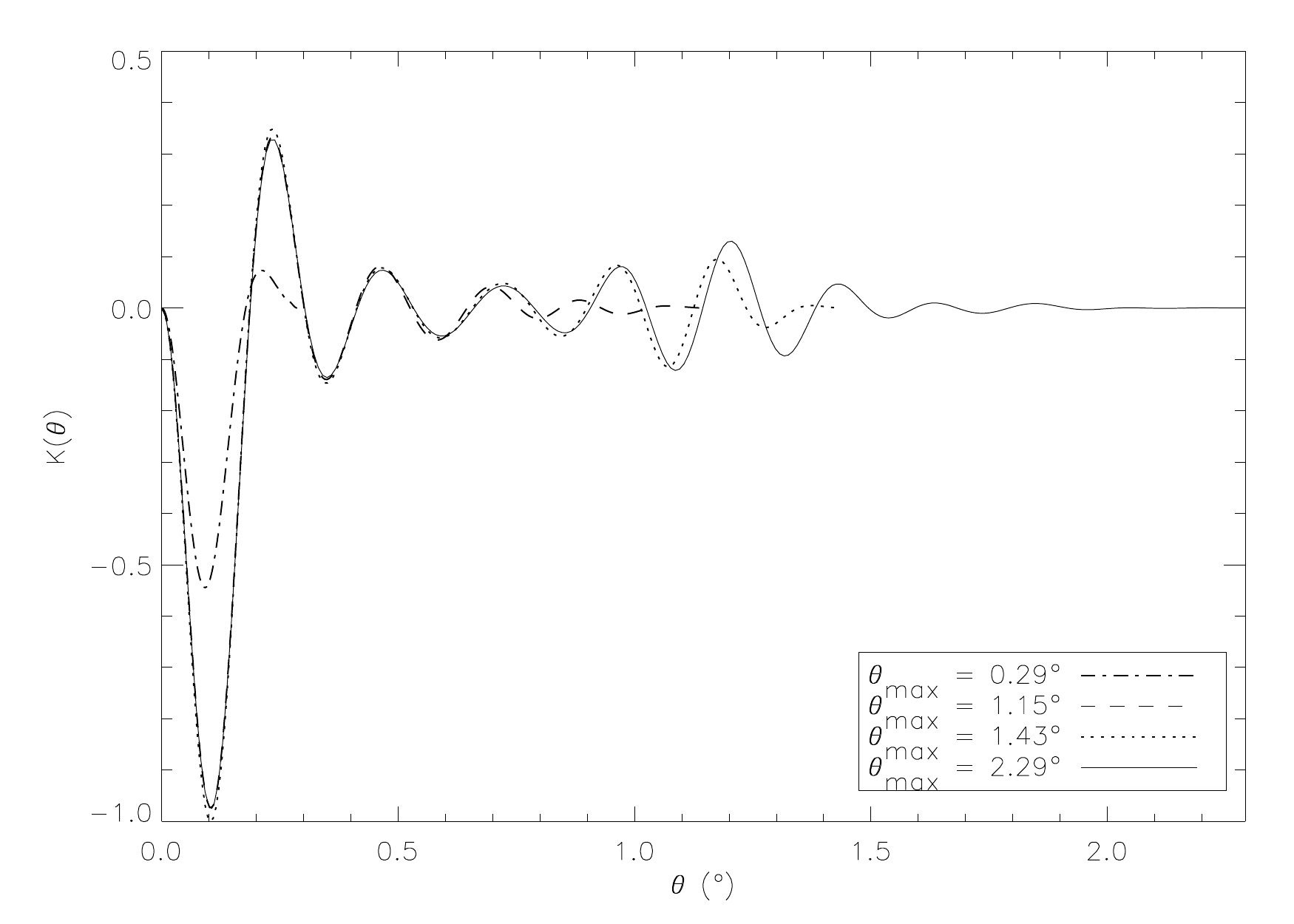}
    \includegraphics[width=5.3cm]{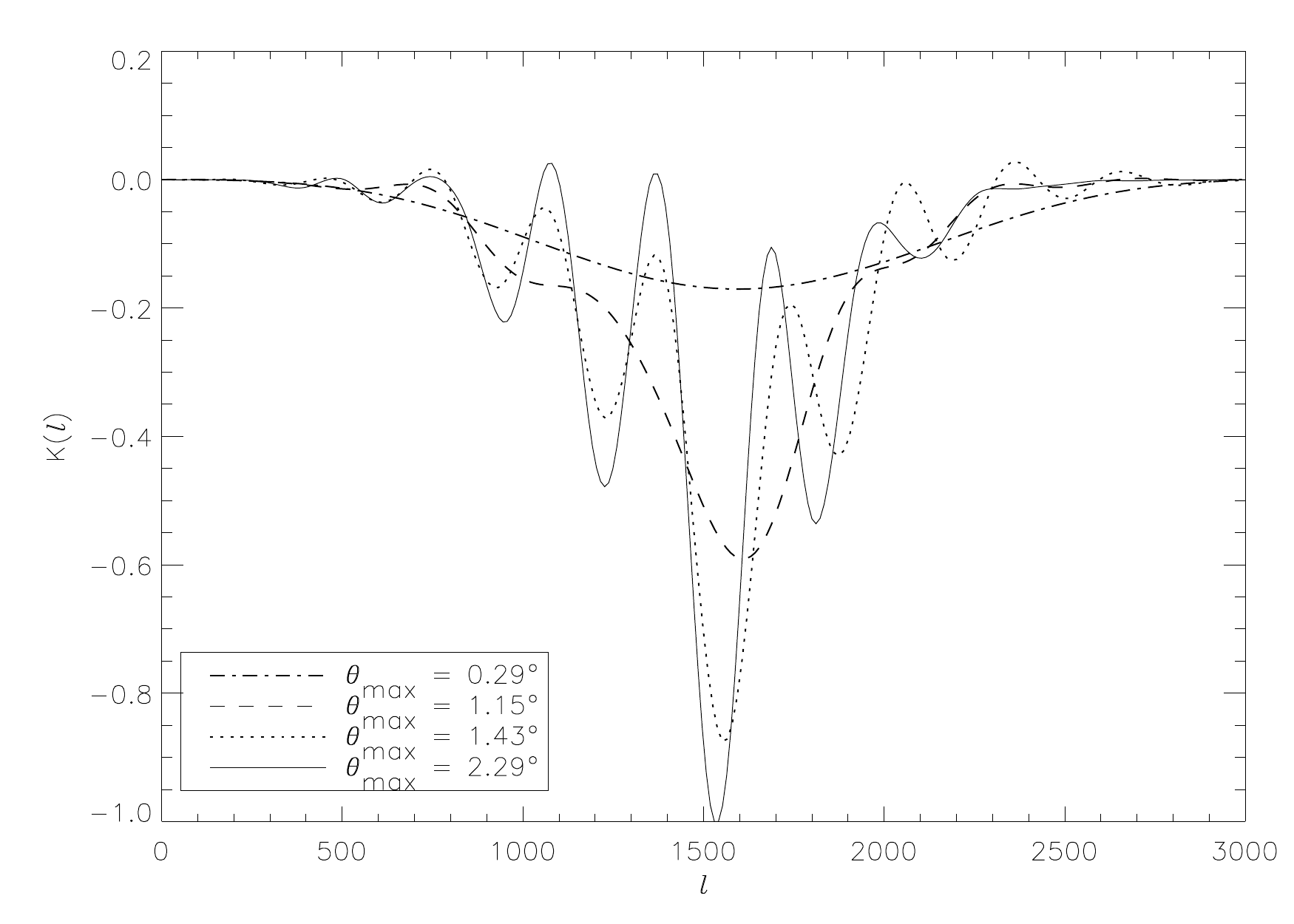}
 \end{center}
\caption{\baselineskip=0.5cm{
{\bf Performance of estimator with truncated angular support.}
We indicate how limiting the angular support of the filter in our
estimator increases the noise.
The panels on the top row refer to the dilatation filter, while the panels on the bottom row refer to the shear filter.
Panel (a) indicates how the estimator variance (with the estimator
normalized to be unbiased) increases
as the angular support (disk radius in degrees) is reduced.  
Panels (b) and (c) indicate the profiles of the optimal truncated estimators
in both angular space and harmonic space. 
}}
\label{Fig:TruncEst}
\end{figure}

In real space the 
minimum variance full-sky estimator kernel for $\kappa _0$ is given by
\ba 
K_{ideal}(\theta )
&=&
\int _0^\infty \ell ~d\ell ~
K_{ideal}(\ell )\cr 
&=&
\int _0^\infty \ell ~d\ell ~
{J_0(\ell \theta )}
\frac{1}{N_{ideal}}
\frac{C(\ell )}
{[C(\ell) +N(\ell )]^2}~
\left[ \frac{d(\ln [C(\ell )])}{d[\ln (\ell )]} +2 \right] .
\ea
This kernel may be inverted using the following inverse Bessel
transform:
\ba 
K_{ideal}(\ell )=\frac{1}{2\pi }\int _0^\infty \theta ~d\theta ~
J_0(\ell \theta )~
K_{ideal}(\theta ). 
\ea
We limit the support of the kernel by requiring that $K(\theta )$ 
be nonzero only for $\theta \le \theta _{max}$ where
$\theta _{max}$ is varied. This is accomplished numerically
by expressing $J_0(\ell \theta )$ as a linear combination
of cubic spline basis functions spanning the interval 
$\theta \in [0, \theta _{max}]$ and optimizing for the shape that minimizes the 
variance calculated according to eqn. (\ref{VVar}). Analogous expressions
may be obtained for the shear by replacing $J_0$ with $J_2.$
Fig. \ref{Fig:TruncEst} shows the variance ratio as a function of $\theta _{max}$ for the dilatation and shear estimators. We observe that the increase in variance at small separations is more modest for the shear estimator.

\section{Results: Dilatation and Shear Reconstruction}

In the previous section we showed how by means
of a linear filter 
${\cal F}_{\kappa _0}$ applied to a temperature map 
$T(\thetab ),$ we may obtain a reconstructed dilatation field
through the product map
\ba
\kappa _{0, rec}(\thetab )
=T(\thetab )({\cal F}_{\kappa_0}T)(\thetab )-c
\ea
where $c$ is a constant offset. We presented
a theoretical derivation of the optimal shape for such a filter.
The best shape for the filter depends both on the cosmological model,
and more importantly on the details of the experiment, because most
of the statistical weight is situated on the smallest angular scales,
near the resolution limit of the experiment. If the primordial anisotropies
dominated to arbitrarily small scales and experiments of unlimited
sensitivity and angular resolution were possible,  
the lensing field could be reconstructed as accurately as desired
simply by increasing the angular resolution and sensitivity.
For a given experiment, the desirable filter shape may be 
intuitively understood as follows. We assume weak lensing (i.e., 
$\vert \kappa \vert \ll 1).$
We want to measure the change in
shape of the power spectrum (or rather of the ``local'' power spectrum 
within a certain finite size patch for the 
case of interest of a non-uniform dilatation). Working in 
the flat sky approximation, which is appropriate because the length 
scales most sensitive to the changes in overall scale are 
small (i.e., $\ell \gg 100 ),$
we define ${\cal C}(\ell )=\ell ^2~C(\ell ).$ Under a small
dilatation, which transforms the temperature map as follows 
\ba
T(\thetab )\to T(\thetab ')=T\Bigl(\exp [-\kappa ]~\thetab \Bigr),
\ea
the power spectrum transforms as
\ba 
{\cal C}(\ell )\to 
{\cal C}\Bigl(\exp [+\kappa ]~\ell \Bigr),
\ea 
so for positive dilatations the power spectrum
in the form ${\cal C}(\ell )$ squeezes to the left in
such a way that the amplitudes of the various features remain
unchanged. 
If the dilatation is small, the change in the power spectrum
is proportional to $\kappa ~{\cal C}'(\ell ).$ 
To make a good filter for implementing the above scheme 
giving the reconstruction with the least noise, we wish to
block modes of wavenumber where ${\cal C}(\ell )$ is approximately
flat, to
allow modes of wavenumber where ${\cal C}(\ell )$ is rising
to pass with a positive phase factor, 
and to
allow modes of wavenumber where ${\cal C}(\ell )$ is falling
to pass with a negative phase factor. Moreover the filter should
combine the different wavenumbers according to inverse variance
weighting in order to minimize the noise of the reconstructed dilatation field.

The quality of the inverse variance reconstruction may be characterized
quantitatively in terms of the variance of the reconstructed field. 
We define a quality factor $Q$ so that 
\ba 
(\delta \kappa )^2\approx \frac {Q}{\cal A}
\ea
where ${\cal A}$ is the area of the region over which the reconstructed
field has been averaged and $(\delta \kappa )$ is
the fluctuation in $\kappa $ averaged over that same area.
Here $Q$ is equal to $\sigma ^2_{\hat \kappa _0}/{\cal A}$ where
$\sigma ^2_{\hat \kappa _0}$ is given in eqn.~(\ref{SoverN}).

Implicit is the assumption
that the noise in the reconstructed dilatation field lacks long
range correlations---that is, it can be treated as white noise 
beyond a certain angular coherence scale. This is indeed the case because 
the filter ${\cal F}_{\kappa _0}$ 
blocks small wave numbers. 
We assume that the area ${\cal A}$ is sufficiently large so that the white
noise regime has been reached. 

\begin{figure}
 \begin{center}
    \includegraphics[width=15cm]{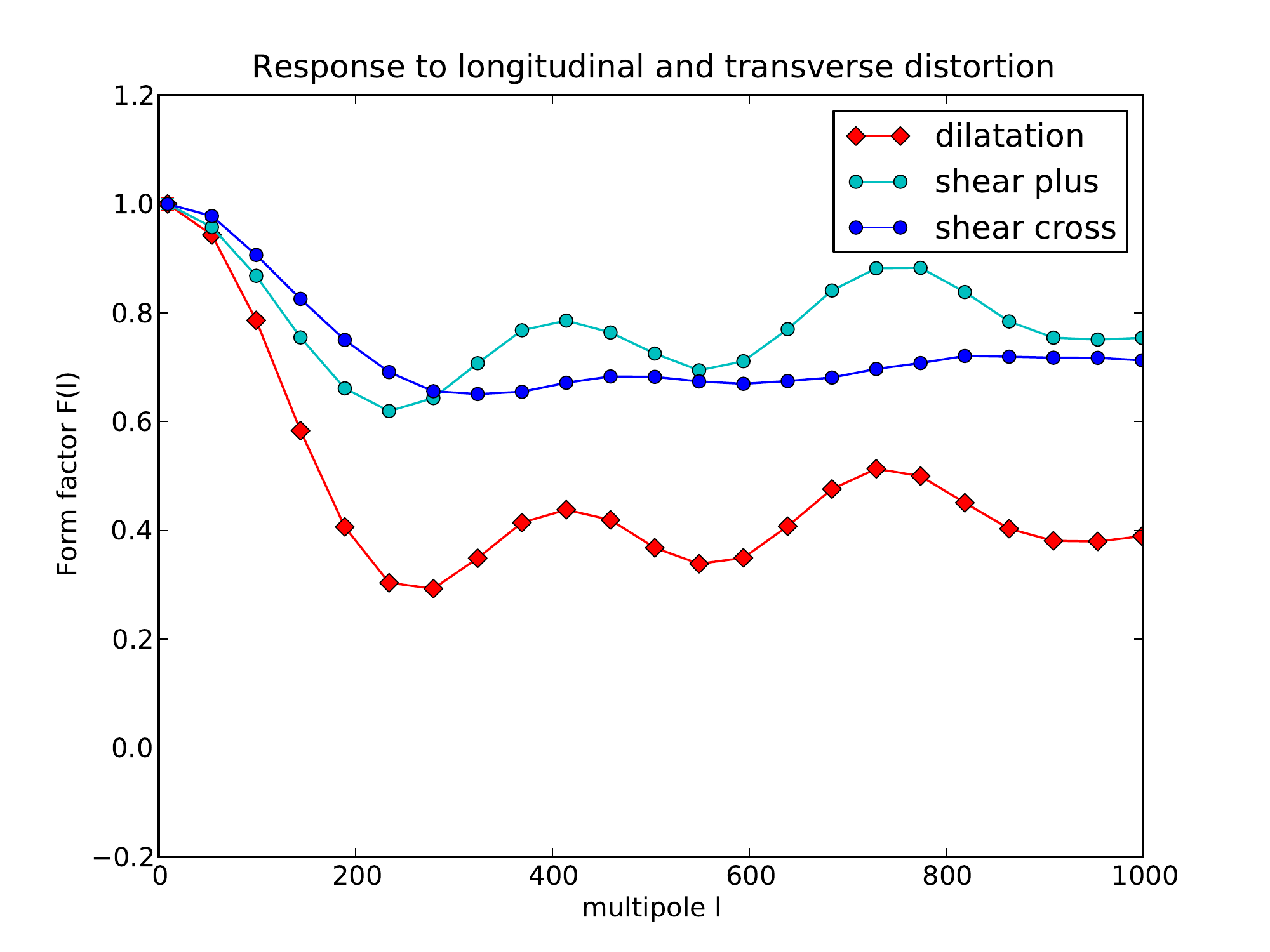}
 \end{center}
\caption{\baselineskip=0.5cm{
{\bf Estimator form factors at large lensing wave numbers.}
We plot
as a function of multipole number
the form factors for the dilatation, longitudinal shear, and cross shear reconstructions using the filter
described in the text.
Noise for the Planck instrument combining the 100+143+217 GHz channels is assumed using bluebook values
and combining channels in the naive way \cite{bluebook}. A deformation field
whose wavenumber constitutes the horizontal axis
was applied to unlensed maps. For the dilatation and the ``plus" shear,
a longitudinal displacement field was used. For the ``cross" shear a transverse deformation was applied.
As a result of symmetry considerations, the responses of the dilatation and plus shear to a transverse
deformation have vanishing expectation values (i.e., they are pure noise) 
and hence are not shown. The same holds for
the cross shear response to a longitudinal deformation.
}}
\label{Fig:FormFactorsConsolidated}
\end{figure}

For a non-uniform dilatation field having a small wavenumber, the previous
analysis carries over with very small corrections. Using the same
filter as for the uniform case, one expects that 
\ba 
\left< \kappa _{0, recon}(\thetab )\right> =\kappa _{0, exact}(\thetab ),
\ea
in other words, that there is negligible bias. However for larger
wavenumbers a form factor appears due to the fact that our estimator
probes the power spectrum over a window of a finite width in angular space, 
and consequently smoothes
the underlying exact dilatation field thus reducing its amplitude. Quantitatively,
if the dilatation field has the form 
\ba
\kappa _0 \cos (\ellb \cdot \thetab +\phi ),
\ea
the expectation value of the reconstructed field takes the form 
\ba 
F(\ell ) \kappa _0 \cos (\ellb \cdot \thetab +\phi ) 
\ea
and as before the noise field will have the same statistics. 
$F(\ell )$ is known as the form factor and satisfies 
$F(0)=1.$ $F(\ell )$ 
falls off for increasing $\ell,$ eventually going to zero.
The form factors are indicated in Fig.~\ref{Fig:FormFactorsConsolidated}.
The shear plus estimator experiences less smoothing than the dilatation estimator because the form of the $\ell$-space filter has less cancellations as can be seen in Fig.~\ref{Fig:TruncEst}.

\begin{figure}
 \begin{center}
   \includegraphics[width=12cm]{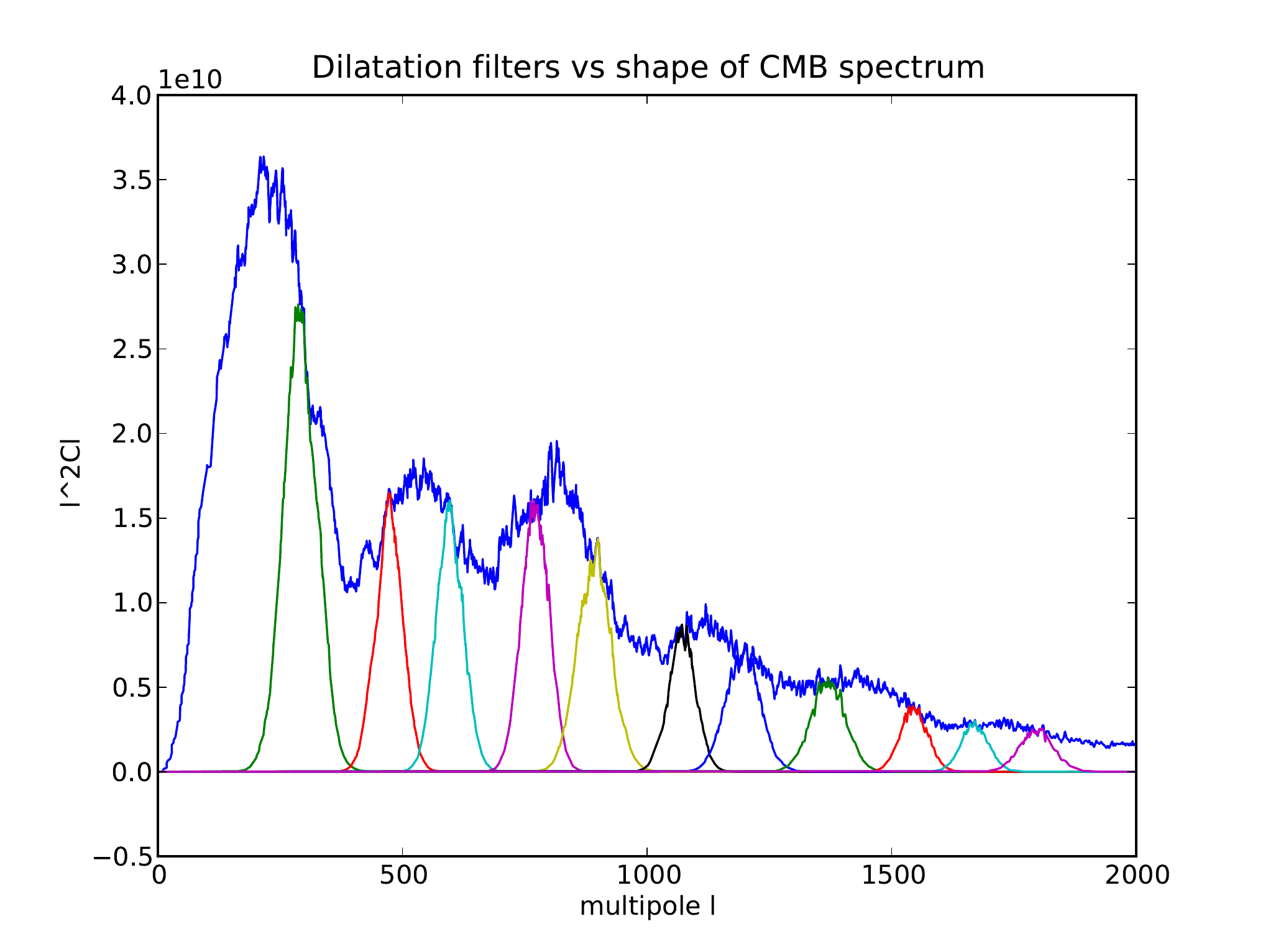}
 \end{center}
 \caption{\baselineskip=0.5cm{
{\bf Alternative filter derivation.}
We generate a dilatation reconstruction filter by an alternate procedure whereby Gaussian filters are placed at the 
locations in multipole space where the magnitude of the derivative of $\ell ^2~C(\ell )$ is greatest. The widths
of the filters are chosen to correspond roughly to the width of the rise and fall of this quantity.
}}
\label{Fig:gaussianfilters}
\end{figure}

In Fig.~\ref{Fig:gaussianfilters} we present an alternative, more heuristic derivation of the dilatation field
reconstruction filter which has the virtue of rendering manifest the contribution of various features in the 
power spectrum to the reconstruction. This alternative filter is a combination of Gaussian filters placed at the locations in multipole space where the magnitude of the derivative of $\ell ^2~C(\ell )$ is greatest, and whose widths
are chosen to correspond roughly to the width of the rise and fall of this quantity.
In Table \ref{SNTable} the $(S/N)^2$ for each of these filters are shown and we indicate their relative statistical weights as well as the $(S/N)^2$ when they are combined using inverse variance weighting. 
Somewhat surprisingly, this naive derivation yields a filter almost as efficient
as the optimal one.

\begin{table*}[htbp]
{\footnotesize 
\begin{center}
\begin{tabular}{|p{3.5cm}|p{3.5cm}|p{3.5cm}|}
\hline
\centering{$\ell _{center}$}  & \centering{$\sigma _\ell$}   & \centering{$\left(S/N\right)^2$ $\deg^{-2}\cdot\kappa_0^{-2}$}  \tabularnewline
\hline \hline
\centering{$300$} &  \centering{50} & \centering{$3.13 \pm 0.26$} \tabularnewline
\hline
\centering{$470$}  & \centering{40}  & \centering{$1.56 \pm 0.13$} \tabularnewline
\hline
\centering{$600$} & \centering{40}  & \centering{$3.02 \pm 0.25$} \tabularnewline
\hline
\centering{$770$}  & \centering{40}  & \centering{$0.13 \pm 0.01$}\tabularnewline
\hline
\centering{$900$} &  \centering{50}  & \centering{$61.51 \pm 5.02$} \tabularnewline
\hline
\centering{$1070$} & \centering{40}  &  \centering{$1.44 \pm 0.12$} \tabularnewline
\hline
\centering{$1200$} & \centering{50}  &  \centering{$30.66 \pm 2.50$} \tabularnewline
\hline
\centering{$1370$} & \centering{50}  &  \centering{$23.24 \pm 1.90$} \tabularnewline
\hline
\centering{$1550$} & \centering{40}  &  \centering{$42.53 \pm 3.47$} \tabularnewline
\hline
 \multicolumn{2}{|c|}{Combined}  &   \centering{$167.22 \pm 13.65$} \tabularnewline
\hline 
 \multicolumn{2}{|c|}{Ideal}  & \centering{$140.50 \pm 11.47$} \tabularnewline
\hline
\end{tabular}
\end{center}
}
\caption{\baselineskip=0.5cm{
{\bf Performance of individual feature filters.}
The first nine entries correspond to Gaussian filters whose centers and widths
are indicated and have been placed where the features of the power spectrum
change the most in response to a small dilatation. The last column expresses
the sensitivity of each filter. The next-to-last row shows $(S/N)^2$
resulting when the filters are combined using inverse covariance 
weighting. This is compared to the ideal filter. The errors
result from Monte Carlo noise.
}}
\label{tab:signaltonoise}
\label{SNTable}
\end{table*}

\begin{figure}
 \begin{center}
    \includegraphics[width=5cm]{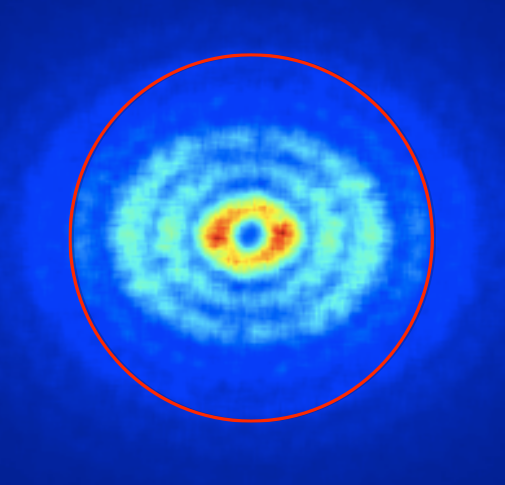}
    \includegraphics[width=5.5cm]{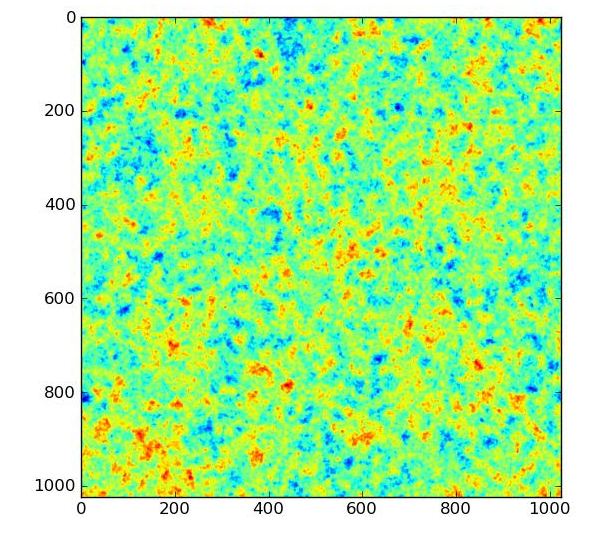}
    \includegraphics[width=5.5cm]{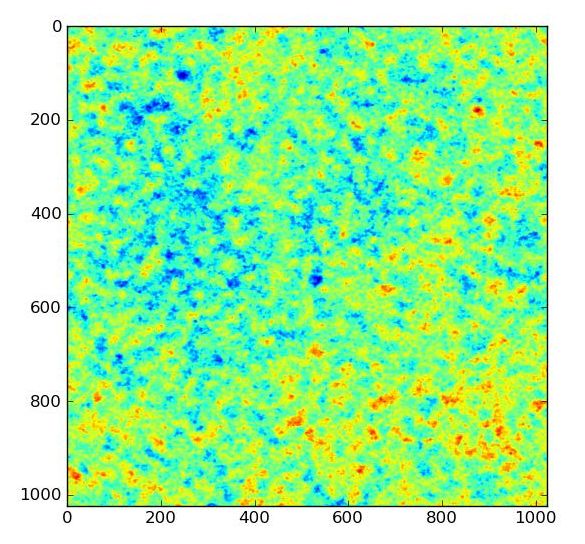}
 \end{center}
\caption{\baselineskip=0.5cm{
{\bf Distortion of power spectrum due to shear.}
In the two-dimensional plot in harmonic space the power
spectrum is shown after a pure shear transformation has
been applied turning the sequences of acoustic peaks
into concentric ellipses rather than concentric circles.
In the real space images (middle, unlensed and right, sheared)
the transformation is hardly apparent to the unaided eye.
}}
\label{FigPowerAniso}
\end{figure}

\begin{figure}[t]
 \begin{center}
    \includegraphics[width=15cm]{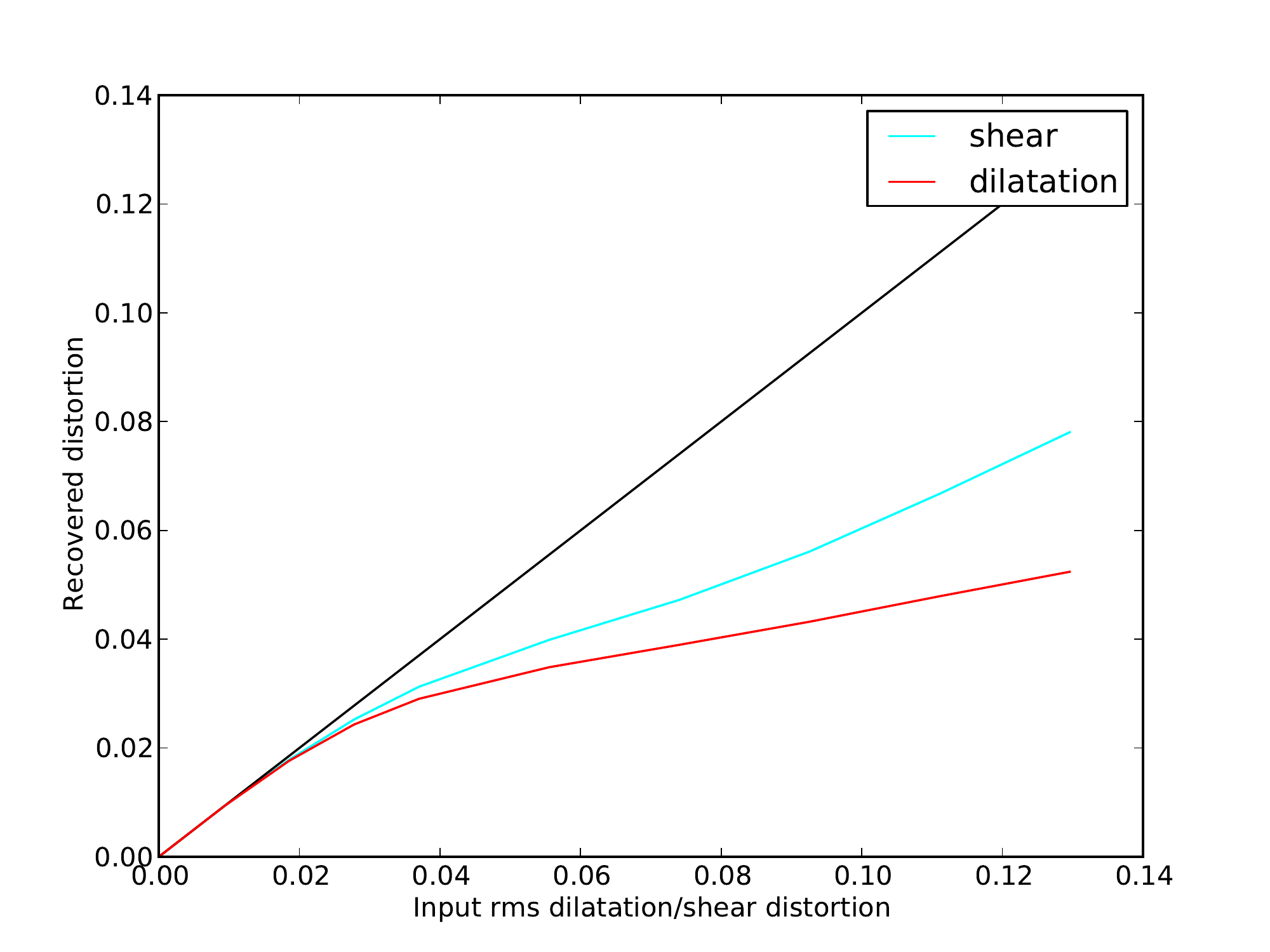}
 \end{center}
\caption{\baselineskip=0.5cm{
{\bf Estimator non-linearity.}
We plot the recovered root-mean-square distortion versus the input
root-mean-square distortion in order to characterize the range
of linear response for our estimator. The input is a long-wavelength 
longitudinal deformation, so that the two-dimension shear and
dilatation are exactly half the one-dimensional dilatation.
Exactly the same non-linearity will plague the quadratic estimator
as well, since the two have been demonstrated to be equivalent
at low wavenumber (as discussed in section \ref{CompQuad}).
}}
\label{Fig:NonLin}
\end{figure}

We now repeat the same analysis for
a constant pure shear deformation, for concreteness
with $\kappa _+>0,$ $\kappa _\times =0,$ there is a stretching
along the $x$-direction with the same amount of compression
along the $y$-direction, so that the temperature map
is deformed according to
\ba 
T(\theta _x,\theta _y)\to 
T'(\theta _x',\theta _y')=
T(e^{-\kappa _+}\theta _x,
e^{+\kappa _+}\theta _y),
\ea
and for the harmonic coefficients one has the following transformation law 
\ba 
T(\ell _x,\ell _y)\to
T'(\ell _x',\ell _y')=
T(e^{+\kappa _+}\ell _x,
e^{-\kappa _+}\ell _y).
\ea
For any power spectrum having a shape like that of the CMB 
temperature, where at all wavenumbers the spectrum is ``red'' 
compared to a white noise spectrum, a pure shear deformation
causes a loss of power for wavevectors oriented close to the stretched principal axis 
and a corresponding increase of power for wavevectors close to the principal axis of compression. 
Along the diagonal direction the power is unaffected.
The estimator developed in the first section of
this paper exploits this anisotropy to reconstruct two components
of the shear field.
The effect is illustrated in Fig. \ref{FigPowerAniso}. In panel (a) 
the acoustic oscillations, which would in the absence of lensing
be visible as a series of
concentric circular rings, have as a result of a pure shear deformation
been deformed into concentric ellipses.
Panels (b) and (c) show simulated temperature maps 
before and after a constant shear deformation, 
here exaggerated in magnitude (with $\kappa _0=0.1$)
for clarity. One observes that the annuli 
become elliptical. Unlike for the dilatation field, where
a scale-invariant spectral index implies that statistically 
the pattern does not change, for pure shear an isotropic
scale invariant random field is deformed into an anisotropic
scale invariant random field, where heuristically one might
say that the series of elliptical (rather than on the average
circular) diffuse motifs have been put down
in a scale-free manner. 
In Fig.~\ref{Fig:NonLin} we show the extent of the region of linear response of the dilatation and shear estimators. Ignoring higher order terms gives rise to a bias which can be corrected \cite{seljak-bis,kesden_2003}.

\section{\label{Combining} Combining the $\kappa _0(\theta ),$ $\kappa _+(\theta )$ and 
$\kappa _\times (\theta )$ reconstructions}

In the previous two sections we have shown how by means of
three filters 
${\cal F}_{\kappa _0},$
${\cal F}_{\kappa _+ },$ and 
${\cal F}_{\kappa _\times }$ we may obtain a noisy
reconstruction of the fields
$\kappa _0      (\thetab ),$
$\kappa _+     (\thetab ),$ and 
$\kappa _\times(\thetab )$ 
in real space. We also saw how the reconstruction was 
essentially finite-range in real space and 
could be made of compact support with very little loss
of information relative to the optimal nonlocal, 
full-sky reconstruction. The reconstructed fields are not independent
but related according to the following relations
\ba
\kappa _0(\ellb ) &=& -\frac{1}{2}(\ell _x^2 +\ell _y^2 )\Phi (\ellb ),\cr 
\kappa _{+}(\ellb ) &=& -\frac{1}{2}(\ell _x^2 -\ell _y^2 )\Phi (\ellb ),\cr 
\kappa _{\times }(\ellb ) &=& -\ell _x\ell _y\Phi (\ellb ),
\ea
which are local in harmonic space.
These consistency relations, expressed in terms of the 
lensing potential $\Phi ,$ can be used to harmonize the 
reconstruction. Even though there are three reconstructed
fields, the cross shear, which would result from a transverse
deformation that cannot be produced by weak lensing, is pure
noise. The noises present in the longitudinal shear and the 
dilatation reconstruction are uncorrelated because one 
involves the $m=0$ sector and the other the $m=\pm 2$ sector.
Harmonization by inverse variance weighting reduces the noise. 
If we consider the noise in the 
reconstruction field to lowest order in the perturbation
expansion (as 4-point functions of the underlying Gaussian
field) ignoring higher-order terms, then there are no 
correlations between reconstructed fields components of 
differing wavenumber. 
Consequently, in order to obtain the lowest noise
estimator of $\Phi ,$ it suffices to consider each
$\ellb $ sector separately and to use the covariance
matrix of the noise in 
$\kappa _0(\ellb ),$
$\kappa _{+}(\ellb ),$ and
$\kappa _{\times }(\ellb )$ 
as a basis for the optimal harmonization. 

\section{Comparison with the quadratic estimator}
\label{CompQuad}

\begin{figure}[t]
\vskip-0.5cm
 \begin{center}
    \includegraphics[width=15cm]{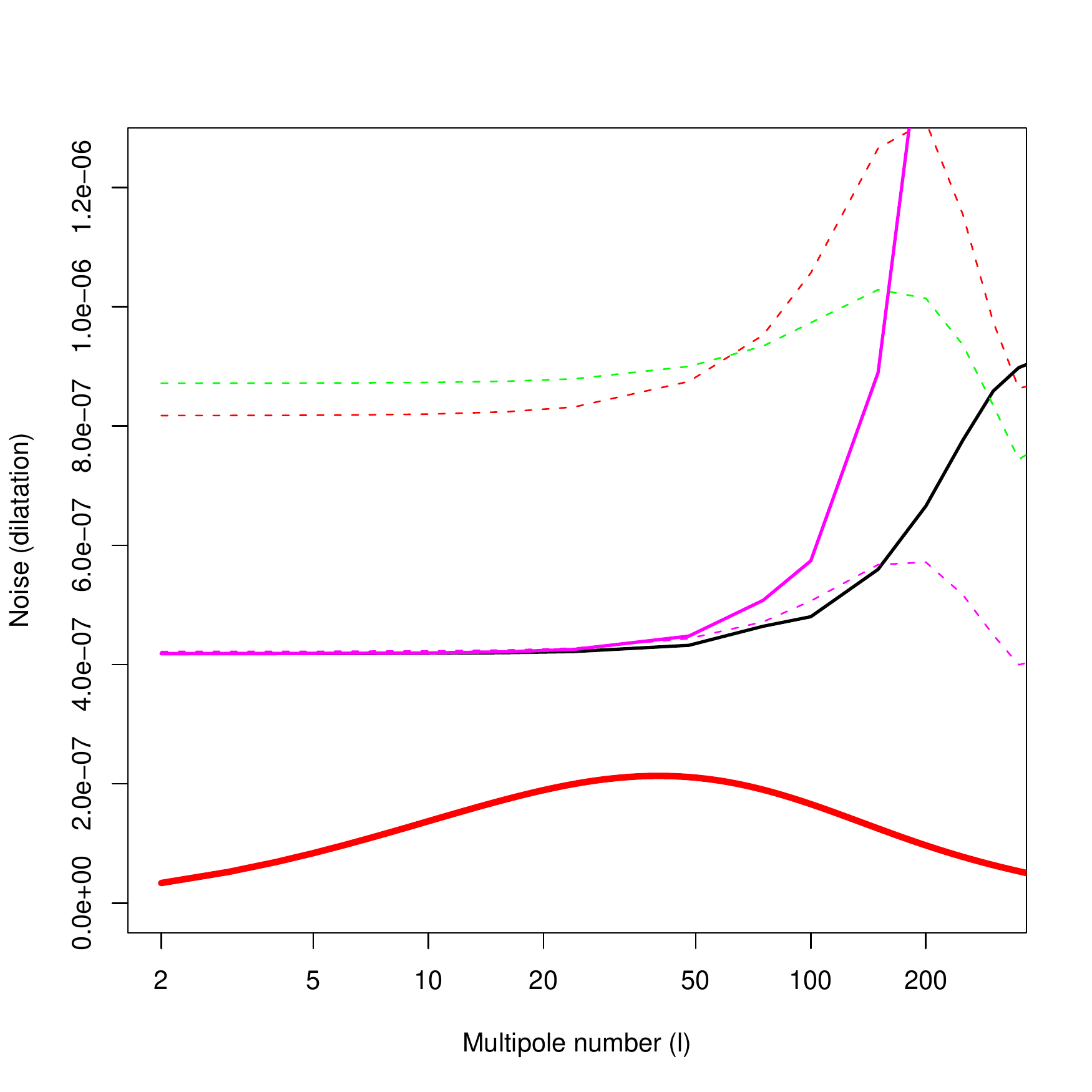}
 \end{center}
 \vskip-0.5cm
\caption{\baselineskip=0.5cm{
{\bf Comparison with quadratic estimator}
We plot the noises of the various estimators
compared to the expected signal (heavy red curve). The quadratic estimator is indicated
in thick black. The dilatation and shear estimators are shown in dashed red and green,
respectively, and when combined nominally give the dashed magenta curve, but when the
imperfect overlap with the expected signal is taken into account, yield the solid
magenta curve. The blue curve would be indicative of the actual noise in the recovered
maps, but if the imperfect overlap where corrected to remove the bias at high 
$\ell $ 
the heavy magenta curve would result.
For comparison we show the predicted lensing signal (as computed by CAMB
for the WMAP best fit model) as the heavy red curve. 
}}
\label{Fig:CompQuadEst}
\end{figure}

In this section we show how the dilatation and shear
estimators developed in this paper 
are related to the linearly optimal
quadratic estimator for the lensing field \cite{hu2001a,hu2001b}. 
We demonstrate that in the low-$\ell $ limit
the quadratic estimator becomes a linear combination
of our dilatation and longitudinal shear
estimators. In fact in this limit the linearly
optimal quadratic estimator
is identical to the inverse variance weighted linear combination
of the
dilatation and longitudinal shear estimators, whose
noises are uncorrelated.
At higher $\ell ,$ the 
approximation used to derive this relation progressively
breaks down. To quantify this degradation, we plot in Fig.
\ref{Fig:CompQuadEst} the 
increase in variance of the combined dilatation
plus longitudinal shear estimators. 
For the reasons already presented concerning the 
minute statistical weight contributed by 
small multipoles, we find it adequate to
work in the flat sky approximation. 

The relation in real space (valid to linear order)
\ba 
\delta T(\thetab )=\nabla \Phi (\thetab )\cdot \nabla T(\thetab )
\ea
in harmonic space translates into 
\ba
\delta T(\ellb )=
\int \frac {d^2\ell '}{(2\pi )^2}
(-\ellb ')\cdot (\ellb -\ellb ')~
\Phi (\ellb ')~
T(\ellb -\ellb ')
\ea
where we define
\ba 
\left< T(\ellb )T(\ellb ') \right> 
=(2\pi )^2 \delta ^2(\ellb +\ellb ')~C(\ell ).
\ea
Since $T_{sky}=T+\delta T,$ it follows that to leading (linear)
order
\ba 
&&
\left< 
T_{sky} \left( \frac{\ellb '}{2}+\ellb \right)
T_{sky} \left( \frac{\ellb '}{2}-\ellb \right)
\right>\cr 
&=&\Phi (\ellb ')\times 
\left[
-\left( \frac{\ellb'}{2}-\ellb \right)\cdot \ellb '~
C\left(\left| \ellb -\frac{\ellb'}{2} \right|\right) 
-\left( \frac{\ellb'}{2}+\ellb \right)\cdot \ellb '~
C\left(\left| \ellb +\frac{\ellb'}{2} \right|\right) 
\right] .
\label{eqn_quad}
\ea
It follows that
\ba 
\hat \Phi (\ellb ')=N^{-1}\sum _{\ellb }
W(\ellb'; \ellb )~
T_{sky} \left( \frac{\ellb '}{2}+\ellb \right)~
T_{sky} \left( \frac{\ellb '}{2}-\ellb \right)
\ea
is the minimum variance unbiased estimator for
the Fourier coefficient $\Phi (\ellb ')$ where
\ba 
W(\ellb '; \ellb )=
\frac{
\left[
-\left( \frac{\ellb'}{2}-\ellb \right)\cdot \ellb '~
C\left(\left| \ellb -\frac{\ellb'}{2} \right|\right)
-\left( \frac{\ellb'}{2}+\ellb \right)\cdot \ellb '~
C\left(\left| \ellb +\frac{\ellb'}{2} \right|\right)
\right]
}{
\left[ 
C\left(\ellb -\frac{\ellb '}{2}\right)+
N\left(\ellb -\frac{\ellb '}{2}\right)
\right]
\left[
C\left(\ellb +\frac{\ellb '}{2}\right)+
N\left(\ellb +\frac{\ellb '}{2}\right)
\right]
}
\label{eqnW}
\ea
and
\ba 
N=\sum _{\ellb } 
\frac{
\left[
-\left( \frac{\ellb'}{2}-\ellb \right)\cdot \ellb '~
C\left(\left| \ellb -\frac{\ellb'}{2} \right|\right)
-\left( \frac{\ellb'}{2}+\ellb \right)\cdot \ellb '~
C\left(\left| \ellb +\frac{\ellb'}{2} \right|\right)
\right] ^2
}{
\left[
C\left(\ellb -\frac{\ellb '}{2}\right)+
N\left(\ellb -\frac{\ellb '}{2}\right)
\right]
\left[
C\left(\ellb +\frac{\ellb '}{2}\right)+
N\left(\ellb +\frac{\ellb '}{2}\right)
\right]
}.
\ea
We may approximate the quantity in the square brackets in eqn.~(\ref{eqn_quad}) 
to linear order to obtain
\ba 
&&
-\left( \frac{\ellb '}{2}-\ellb \right)\cdot \ellb ' ~C\left( \left| \ellb -\frac{\ellb '}{2}\right| \right)
-\left( \frac{\ellb '}{2}+\ellb \right)\cdot \ellb ' ~C\left( \left| \ellb +\frac{\ellb '}{2}\right| \right) \cr 
&=&-{\ell '}^2~C(\ell )-(\ellb \cdot \ellb ')^2\frac{1}{\ell }
\frac{\partial C(\ell )}{\partial \ell }\cr  
&=&-\frac{1}{2}\frac{{\ell '}^2}{\ell ^2}\frac{\partial [\ell ^2C(\ell )]}{\partial \ln [\ell ]}
-
\left( \frac{
(\ellb \cdot \ellb ')^2-\frac{1}{2}{\ell '}^2\ell ^2
}{\ell ^2}\right) 
\frac{\partial [C(\ell )]}{\partial \ln [\ell ]}\cr 
&=&-\frac{1}{2}{\ell '}^2\left\{  
\frac{1}{\ell ^2}
\frac{\partial [\ell ^2C(\ell )]}{\partial \ln [\ell ]}
+\cos [2\Theta ]
\frac{\partial [C(\ell )]}{\partial \ln [\ell ]}
\right\} 
\label{eqnN}
\ea
where $\Theta $ is the angle between $\ellb $ and $\ellb '.$ 
The expression is remarkably similar to a linear combination
of the expressions appearing in the dilatation and the pure shear 
reconstruction of the previous section. 

Approximating the numerator of (\ref{eqnW}) as (\ref{eqnN})
and the denominator of (\ref{eqnW}) as 
$[ C(\ell )+ N(\ell )]^2,$ we may write the 
three estimators in the following unified manner:
\ba 
\hat D(\ellb ')&=&
\frac{1}{2}
\frac{\cal A}{N_D}
\int \frac{d^2\ell }{(2\pi )^2}
\frac{
{[C(\ell )]}
}
{[C(\ell )+N(\ell )]^2 
}
\left[
\frac{\partial (\ln [\ell ^2C(\ell )])}{\partial [\ln (\ell )]}
\right] ~
T_{sky}\left( \frac{\ellb '}{2}+\ellb \right) ~
T_{sky}\left( \frac{\ellb '}{2}-\ellb \right)\cr
\hat S_L(\ellb ')&=&
\frac{1}{2}
\frac{\cal A}{N_{S_L}}
\int \frac{d^2\ell }{(2\pi )^2}
\frac{
{[C(\ell )]}
}
{[C(\ell )+N(\ell )]^2       
}
\left[
\cos [2\Theta (\ellb ,\ellb ')]
\frac{\partial (\ln [C(\ell )])}{\partial [\ln (\ell )]}
\right] ~
T_{sky}\left( \frac{\ellb '}{2}+\ellb \right) ~
T_{sky}\left( \frac{\ellb '}{2}-\ellb \right)\cr
\hat Q(\ellb ')&=&
\frac{1}{2}
\frac{\cal A}{N_{Q}}
\int \frac{d^2\ell }{(2\pi )^2}
\frac{
{[C(\ell )]}
}
{[C(\ell )+N(\ell )]^2
} 
\left[
\frac{\partial (\ln [\ell ^2C(\ell )])}{\partial [\ln (\ell )]}
+
\cos [2\Theta (\ellb ,\ellb ')]
\frac{\partial (\ln [C(\ell )])}{\partial [\ln (\ell )]}
\right] \cr 
&&\qquad \times
T_{sky}\left( \frac{\ellb '}{2}+\ellb \right) ~
T_{sky}\left( \frac{\ellb '}{2}-\ellb \right)
\ea
where
\ba
N_D(\ellb ')&=&
\frac{1}{2}
{\cal A}
\int \frac{d^2\ell }{(2\pi )^2}
\frac{
{[C(\ell )]}^2
}
{[C(\ell )+N(\ell )]^2
}
\left[
\frac{\partial (\ln [\ell ^2C(\ell )])}{\partial [\ln (\ell )]}
\right] ^2\cr 
N_{S_L}(\ellb ')&=&
\frac{1}{2}
{\cal A}
\int \frac{d^2\ell }{(2\pi )^2}
\frac{
{[C(\ell )]}^2
}
{[C(\ell )+N(\ell )]^2
}
\left[
\cos [2\Theta (\ellb ,\ellb ')]
\frac{\partial (\ln [C(\ell )])}{\partial [\ln (\ell )]}
\right] ^2\cr
N_Q(\ellb ')&=&
\frac{1}{2}
{\cal A}
\int \frac{d^2\ell }{(2\pi )^2}
\frac{
{[C(\ell )]}^2
}
{[C(\ell )+N(\ell )]^2
}
\left[
\frac{\partial (\ln [\ell ^2C(\ell )])}{\partial [\ln (\ell )]}
+
\cos [2\Theta (\ellb ,\ellb ')]
\frac{\partial (\ln [C(\ell )])}{\partial [\ln (\ell )]}
\right] ^2 
\ea
and $N_Q=N_D+N_{S_L}$ because the cross term cancels under the integral.

The quantities $N_D,$ $N_{S_L},$ $N_Q$ express the amount of information provided
by the respective estimators, and we have shown that in the limit where $\ellb '$
is small 
\ba 
\hat Q(\ellb ')=
\frac{N_D(\ell ')} {N_D(\ell ')+N_{S_L}(\ell ')}
\hat D(\ellb ')
+
\frac{N_{S_L}(\ell ')} {N_D(\ell ')+N_{S_L}(\ell ')}
\hat S_L(\ellb ').
\ea

The exact expressions in Fourier space for the dilatation and shear statistics at not necessarily small
wavenumber $\ellb '$ are
\ba
\hat D(\ellb ')&=&\frac{{\cal A}}{2}{1\over N_D(\ell ')}
\int \frac{d^2\ell }{(2\pi )^2}
\left\{ 
F\left(\frac{\ellb '}{2}+\ellb \right) +
F\left(\frac{\ellb '}{2}-\ellb \right)
\right\}~
T_{sky}\left( \frac{\ellb '}{2}+\ellb \right) ~
T_{sky}\left( \frac{\ellb '}{2}-\ellb \right)~\cr
\hat S_L(\ellb ')&=&\frac{{\cal A}}{2}{1 \over N_{S_L}(\ell ')}
\int \frac{d^2\ell }{(2\pi )^2}
\biggl\{
G\left(\frac{\ellb '}{2}+\ellb \right) 
\cos \left[ 2\Theta \left(  \frac{\ellb '}{2}+\ellb , \ellb '\right) \right]\cr
&&\qquad\qquad+
G\left(\frac{\ellb '}{2}-\ellb \right)
\cos \left[ 2\Theta \left(  \frac{\ellb '}{2}-\ellb , \ellb '\right) \right]
\biggr\}~
T_{sky}\left( \frac{\ellb '}{2}+\ellb \right) ~
T_{sky}\left( \frac{\ellb '}{2}-\ellb \right)~
\ea
where
\ba 
&&F(\bar \ell )=
\frac{C(\bar \ell )}
{[C(\bar \ell )+N(\bar \ell )]^2}
\frac{\partial ( \ln [\bar \ell ^2C(\bar \ell)]) }
{\partial (\ln [\bar \ell ]) },\cr 
&&G(\bar \ell )=
\frac{C(\bar \ell )}
{[C(\bar \ell )+N(\bar \ell )]^2}
\frac{\partial ( \ln [C(\bar \ell)]) }
{\partial (\ln [\bar \ell ]) }.
\ea
To express compactly the increase in variance due to the non-optimality for $\ellb '$
away from zero, it is useful to define the inner product 
\ba 
\left< \hat A, \hat B\right> = 
\frac{{\cal A}}{2}\int \frac{d^2\ell }{(2\pi )^2}
(C+N)\left( \frac{\ellb '}{2}+\ellb \right)
(C+N)\left( \frac{\ellb '}{2}-\ellb \right)
A(\ellb )B(\ellb )
\ea 
where
\ba
\hat A = 
\frac{{\cal A}}{2}{1\over N_A}
\int \frac{d^2\ell }{(2\pi )^2}~A(\ellb )~
T_{sky}\left( \frac{\ellb '}{2}+\ellb \right) ~
T_{sky}\left( \frac{\ellb '}{2}-\ellb \right)~
\ea 
and $\hat B$ is defined analogously. 

In this notation the various estimators have the following shape functions:
\ba 
Q(\ellb ;\ellb ')&=&
\frac{2}{(\ell ')^2}
\frac{
-\left( \frac{\ellb'}{2}-\ellb \right)\cdot \ellb '~
C\left(\left| \ellb -\frac{\ellb'}{2} \right|\right)
-\left( \frac{\ellb'}{2}+\ellb \right)\cdot \ellb '~
C\left(\left| \ellb +\frac{\ellb'}{2} \right|\right)
}{
(C+N)\left( \frac{\ellb '}{2}+\ellb \right)
(C+N)\left( \frac{\ellb '}{2}-\ellb \right)
}\cr
D(\ellb , \ellb ')&=&
\frac{1}{2}
\left\{
F\left(\frac{\ellb '}{2}+\ellb \right) +
F\left(\frac{\ellb '}{2}-\ellb \right)
\right\},\cr 
S_L(\ellb , \ellb ')&=&
\frac{1}{2}
\Biggl\{
G\left(\frac{\ellb '}{2}+\ellb \right)
\cos \left[ 2\Theta \left(  \frac{\ellb '}{2}+\ellb , \ellb '\right) \right]
\cr
&&\qquad +
G\left(\frac{\ellb '}{2}-\ellb \right)
\cos \left[ 2\Theta \left(  \frac{\ellb '}{2}-\ellb , \ellb '\right) \right]
\Biggr\} .
\ea
It follows that the increase in the variance is given by the 
following expression for the secant squared
\ba 
\frac{
\sigma ^2(D)
}{
\sigma ^2(Q)
}=
\frac{
\left< \hat Q, \hat Q\right> 
\left< \hat D, \hat D\right> 
}{
\left< \hat Q, \hat D\right> ^2} .
\ea

In Fig.~\ref{Fig:CompQuadEst} we indicate the performance of the several estimators at finite $\ell '.$  
We plot 
$C_{\ell}^{(\delta \kappa_0)(\delta \kappa _0)}$(i.e. the expectation value of the square modulus of the harmonic coefficient on the sphere
for the reconstruction noise) for the estimators $Q,$ $D,$ and $S_L,$ in black, dashed red and dashed green, respectively. The dashed magenta curve indicates the noise obtained by combining
$D$ and $S_L$ using inverse variance weighting. The solid magenta curve indicates the correction when the lack of overlap is taken into account to render the estimator unbiased at large wavenumber. We assume an instrument noise
combining the Planck 100+143+217 GHz channels according to the specifications
given in the bluebook \cite{bluebook}. 
At low wavenumber, $Q$ exhibits a flat (i.e., white noise)
spectrum, which subsequently divergences at large $\ell '.$ At very low $\ell '$
we observe that the noise from $Q$ is the same as the noise from $D$ and $S_L$
combined in quadrature, as shown theoretically in the text, but at higher
$\ell '$ the approximations used break down and a noise excess is observed. 
We observe that for $\ell \ltorder 70,$ the difference in performance between
the estimator developed here and the linearly optimal quadratic
estimator is minimal. At higher wavenumbers beyond $\ell \gtorder
100 ,$ however, the variance increases rapidly due to lack of overlap with the ideal
kernel. There is a priori no reason why a real space approach
could not be extended to higher wavenumbers for the lensing field.
However, in the present paper we do not explore how this would work. 

\section{Concluding remarks}

We have demonstrated how to reconstruct in real
space using a filter of compact support the 
weak gravitational lensing field, here represented 
as three fields, a dilatation field $\kappa _0(\thetab )$ and
the two components of the pure shear distortion field
$\kappa _+(\thetab )$ and $\kappa _\times (\thetab ).$
The three fields are related by 
a set of nonlocal consistency conditions, which may subsequently
be exploited to reduce the noise of the reconstruction.
Except for an integration constant
and two translational and one rotational zero modes, the weak lensing may alternatively
and equivalently 
be described by either (1) a gravitational lensing potential
$\Phi (\thetab ),$ (2) a displacement field ${\boldsymbol {\xi}}(\thetab ),$
$\Phi (\thetab ),$ (2) a displacement field ${\boldsymbol{\xi}(\thetab )},$
or (3) the dilatation field $\kappa _0(\thetab )$ and
the two components of the pure shear distortion field
$\kappa _+(\thetab )$ and $\kappa _\times (\thetab ).$
In this paper we argue that for the purpose of reconstruction
the representation (3) is advantageous because this is the 
representation for which the lensing field bears a local relation
to the real space CMB maps. This locality comes at a price
because the three components are not independent and subject
to nonlocal consistency conditions, which may be exploited 
to improve the reconstruction. 
Locality allows different regions of the sky to be analyzed
independently in a natural way, quite unlike the quadratic 
reconstruction in harmonic space, where the entire sky must
be analyzed simultaneously. This approach and variations thereof
hold promise for dealing with partial sky coverage and
excised point sources. 
For the filters developed in this paper there is very little loss
of information for the lensing field at low wavenumbers. However
at larger wavenumbers the lensing signal is attenuated according
to a wavenumber dependent form factor, which can be deconvolved
by applying a correction filter.

\vspace{0.6cm}
\noindent
\textbf{Acknowledgements:}
MB and KM acknowledge support from a joint CNRS/NRF travel grant.
The work of MB, CSC and MR was supported in part by the Projet
Blanc VIMS-PLANCK of the Agence Nationale de la Recherche. 
KM and CSC are supported by the South African National Research
Foundation.

\end{document}